\documentclass[preprint,11pt,authoryear]{elsarticle}

\usepackage{mathtools}
\usepackage{amsmath, amssymb}
\usepackage{enumitem}
\usepackage{mathrsfs}
\usepackage{listings}
\usepackage[hidelinks]{hyperref}%
\usepackage{graphicx}%
\usepackage{dcolumn}%
\usepackage{bm}%

\usepackage{soul}       %
\usepackage{fixmath}
\usepackage{xcolor}

\definecolor{boxblue}{RGB}{238,238,255}
\definecolor{boxblueborder}{RGB}{170,170,255}
\definecolor{boxgrey}{RGB}{238,238,238}
\definecolor{boxgreyborder}{RGB}{191,191,191}

\definecolor{redcol}{RGB}{200,30,30}
\definecolor{bluecol}{RGB}{30,30,200}
\definecolor{greencol}{RGB}{30,150,30}
\definecolor{nullcol}{RGB}{240,240,240}

\usepackage{amsthm}
\usepackage[ruled,linesnumbered]{algorithm2e}
\SetKwProg{Fn}{Function}{}{end}

\usepackage{mwe}
\usepackage{braket}      %
\usepackage{subcaption}

\usepackage{tikz}
\usetikzlibrary{
    graphs,
    shapes.geometric,
    arrows.meta,
    bending,
    positioning,
    knots,
    intersections,
    calc,
    math,
    angles,
    quotes}
\usepackage[precision=2, unit=mm]{lengthconvert} %
\usetikzlibrary{patterns}

\usepackage{booktabs}
\usepackage{threeparttable}
\usepackage{xurl}%

\usepackage{setspace}
\usepackage[margin=1in]{geometry}

\newcommand{\actset}{V}                  %

\newcommand{\alphabet}{\mathcal{L}}      %
\newcommand{\lin}[1]{\mathrm{Lin}(#1)}   %

\newcommand{\smat}{\boldsymbol{P}}       %
\newcommand{\blockset}{\mathcal{B}}      %
\newcommand{\tile}{\mathrm{tile}}        %
\newcommand{\bc}{\mathrm{BC}}            %
\newcommand{\kstar}{k^{\ast}}            %

\newcommand{\gs}{\gamma_1}               %
\newcommand{\gt}{\gamma_2}               %

\newtheorem{definition}{Definition}[section] %
\newtheorem{theorem}{Theorem}[section]       %
\newtheorem{assumption}{Assumption}[section]
\newtheorem{example}{Example}[section]

\newtheorem{proposition}{Proposition}[section]
\newtheorem{corollary}{Corollary}[section]
\newtheorem{lemma}{Lemma}[section]

\theoremstyle{remark}

\newcommand{\Nset}{\mathcal{N}}            %
\newcommand{\modelspace}{\mathbb{R}^{\Nset}} %
\newcommand{\simplexN}{\Delta(\Nset)}       %

\newcommand{\rN}{r_{\Nset}}                 %
\newcommand{\rB}{r_{\blockset}}             %
\newcommand{\rT}{r_{\mathcal{T}}}           %
\newcommand{\rP}{r_{\mathcal{P}}}           %
\newcommand{\rPi}{r_{\Pi}}                  %
\newcommand{\rF}{r_{\mathrm{F}}}            %
\newcommand{\rL}{r_{\alphabet}}             %
\newcommand{\Phik}[1][k]{\Phi_{#1}}         %
\newcommand{\ctree}{\mathcal{K}}            %

\newcommand{\dfgmap}{\mathrm{dfg}}
\newcommand{\seqmap}{\mathrm{seq}}

\DeclarePairedDelimiter{\trace}{\langle}{\rangle}

\newcommand{\linrel}[1][D]{\kappa_{#1}}     %

\newcommand{\dBA}{d_{\mathrm{BA}}}          %
\newcommand{\dSMD}{d_{\mathrm{SMD}}}        %
\newcommand{\dT}{d_{\mathcal{T}}}           %
\newcommand{\dP}{d_{\mathcal{P}}}           %

\newcommand{\mBi}{m_B^{\mathrm{id}}}        %
\newcommand{\mBe}{m_B^{\mathrm{ex}}}        %

\begin{document}

\begin{frontmatter}

\title{Resolution limits for process comparison from event data}

\author[bham]{Antony R. Lee\corref{cor1}}
\ead{arl290@student.bham.ac.uk}
\cortext[cor1]{Corresponding author.}
\author[bham]{Peter Ti\v{n}o}
\author[qub]{Iain B. Styles}

\affiliation[bham]{organization={School of Computer Science, University of Birmingham}, city={Birmingham}, postcode={B15 2TT}, country={United Kingdom}}
\affiliation[qub]{organization={School of Electronics, Electrical Engineering and Computer Science, Queen's University Belfast}, city={Belfast}, postcode={BT7 1NN}, country={United Kingdom}}

\begin{abstract}
One hospital runs bloods and imaging at the same time. Another runs them one after the other, in either order, equally often. Knowing which actually happened, and how it is recorded in data, is critical for all operational managers. In process mining, the standard approach is to construct an event log, and attempt to discover concurrent and sequential processes in a data-driven way. We show this standard approach, built on the stochastic language of an event log, reports only the assumptions of its discovery algorithm, because every such log is explained equally well by a model with no concurrency at all. Further, before any data is acquired, we characterise when data can and cannot distinguish concurrent behaviour. Where it cannot, the distinction is recoverable from evidence the stochastic language discards, such as the times at which activities start and end, or object-centric records that fix an order within an execution. The remedy is therefore a choice of what is recorded, rather than a larger sample. This impacts decision making, as planning resource for truly concurrent services is very different from sequential services.
\end{abstract}


\end{frontmatter}

\section{Introduction}
\label{sec:intro}

One hospital runs bloods and imaging at the same time. Another runs them one after the other, in one order as often as the other. An event log records each execution of the process as a sequence, so both write the same activities into every record in the same proportions, and a comparison of the two logs returns them as the same pathway (Figure~\ref{fig:procedure}(a)). A quality-improvement lead comparing several hospitals, an operations manager auditing one site after a change and a commissioner benchmarking two providers all decide on the strength of such a comparison \citep{belienFiftyYearsOperational2025,boyleFrameworkDevelopingGeneralisable2022,fattahiResourcePlanningStrategies2023,Esensoy_2018}, and process mining supplies the pathways \citep{mazharStochasticAwareComparativeProcess2023,georgievComparingCarePathways2025}. Only the first hospital needs a patient and a scanner at the same moment, and a capacity plan depends on which of the two a site is running.

\providecolor{boxblue}{RGB}{238,238,255}
\providecolor{boxblueborder}{RGB}{170,170,255}
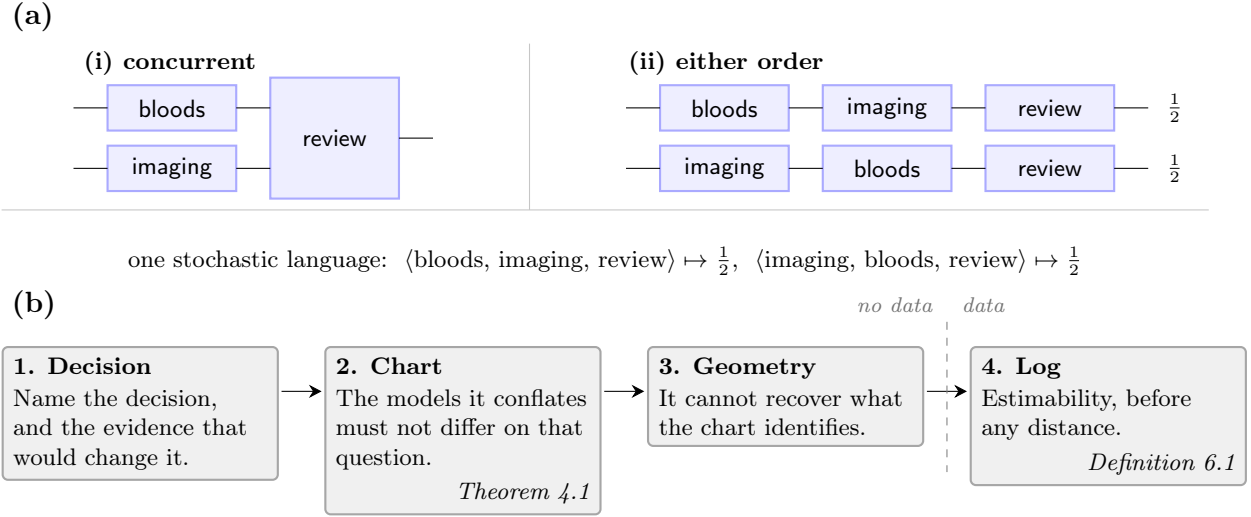
\begin{figure*}[t]
\centering
\begin{tikzpicture}[
  >={Stealth[width=5pt]},
  font=\footnotesize,
  act/.style={draw=boxblueborder, thick, fill=boxblue, minimum width=17mm,
              minimum height=6mm, font=\footnotesize\sffamily, outer sep=0pt},
  wire/.style={-},
  wt/.style={font=\footnotesize},
  ttl/.style={font=\footnotesize\bfseries, anchor=west},
  stage/.style={draw=black!35, thick, fill=black!6, rounded corners=2pt,
                align=left, inner sep=4pt,
                text width={(\textwidth-88pt)/4}, anchor=north west},
  dstage/.style={stage},
  ar/.style={->, shorten >=1pt, shorten <=1pt},
]

\node[font=\bfseries, anchor=west] at (-0.35,4.95) {(a)};

\node[ttl] at (0.60,4.35) {(i) concurrent};
\node[act] (b1) at (1.90,3.75) {bloods};
\node[act] (i1) at (1.90,2.95) {imaging};
\node[act, minimum height=1.6cm] (v1) at (4.05,3.35) {review};
\draw[wire] (0.60,3.75) -- (b1);
\draw[wire] (0.60,2.95) -- (i1);
\draw[wire] (b1.east) -- ($(v1.south west)!0.75!(v1.north west)$);
\draw[wire] (i1.east) -- ($(v1.south west)!0.25!(v1.north west)$);
\draw[wire] (v1.east) -- (5.35,3.35);

\draw[gray!45, thin] (6.625,2.40) -- (6.625,4.60);

\node[ttl] at (7.80,4.35) {(ii) either order};
\node[act] (b2) at (9.20,3.75) {bloods};
\node[act] (i2) at (11.35,3.75) {imaging};
\node[act] (v2) at (13.50,3.75) {review};
\draw[wire] (7.90,3.75) -- (b2); \draw[wire] (b2) -- (i2);
\draw[wire] (i2) -- (v2);        \draw[wire] (v2) -- (14.80,3.75);
\node[wt] at (15.15,3.75) {$\tfrac12$};

\node[act] (i3) at (9.20,2.95) {imaging};
\node[act] (b3) at (11.35,2.95) {bloods};
\node[act] (v3) at (13.50,2.95) {review};
\draw[wire] (7.90,2.95) -- (i3); \draw[wire] (i3) -- (b3);
\draw[wire] (b3) -- (v3);        \draw[wire] (v3) -- (14.80,2.95);
\node[wt] at (15.15,2.95) {$\tfrac12$};

\draw[gray!45, thin] (-0.35,2.40) -- (15.60,2.40);
\node[font=\footnotesize, anchor=north] at (7.625,2.05)
  {one stochastic language:\ \ $\langle$bloods, imaging, review$\rangle\mapsto\tfrac12$,\ \
   $\langle$imaging, bloods, review$\rangle\mapsto\tfrac12$};

\node[font=\bfseries, anchor=west] at (-0.35,1.15) {(b)};

\node[stage] (s1) at (-0.35,0.60)
  {\textbf{1. Decision}\\[1pt] Name the decision, and the evidence that would change it.};
\node[stage, right=6mm of s1.north east, anchor=north west] (s2)
  {\textbf{2. Chart}\\[1pt] The models it conflates must not differ on that question.\\[2pt]
   \hspace*{\fill}\textit{Theorem~\ref{thm:kerL-linrel}}};
\node[stage, right=6mm of s2.north east, anchor=north west] (s3)
  {\textbf{3. Geometry}\\[1pt] It cannot recover what the chart identifies.};
\node[dstage, right=6mm of s3.north east, anchor=north west] (s4)
  {\textbf{4. Log}\\[1pt] Estimability, before any distance.\\[2pt]
   \hspace*{\fill}\textit{Definition~\ref{def:estimability}}};

\draw[ar] ([yshift=-6mm]s1.north east) -- ([yshift=-6mm]s2.north west);
\draw[ar] ([yshift=-6mm]s2.north east) -- ([yshift=-6mm]s3.north west);
\draw[ar] ([yshift=-6mm]s3.north east) -- ([yshift=-6mm]s4.north west);

\draw[dashed, black!50] ($(s3.north east)!0.5!(s4.north west)+(0,0.32)$)
   -- ($(s3.south east)!0.5!(s4.north west)+(0,-1.0)$);
\node[anchor=south east, font=\scriptsize\itshape, black!55]
   at ($(s3.north east)!0.5!(s4.north west)+(-0.06,0.32)$) {no data};
\node[anchor=south west, font=\scriptsize\itshape, black!55]
   at ($(s3.north east)!0.5!(s4.north west)+(0.06,0.32)$) {data};
\end{tikzpicture}
\caption{The conflation, and the procedure it forces. (a) Pathway (i) runs bloods and imaging concurrently, (ii) runs them one at a time, each order half the time. The two therefore emit one stochastic language, so no comparison computed from traces separates the pair (Corollary~\ref{cor:lin-boundary}), and only (i) requires a patient and a scanner at the same time. (b) Each stage constrains the one after it, so the least detailed adequate chart is settled before any data is acquired. Stage~2's feasibility test is Theorem~\ref{thm:kerL-linrel}, and Stage~4, beyond the dashed boundary and the only stage requiring data, certifies the log by the estimability of Definition~\ref{def:estimability}, which reports one way Proposition~\ref{prop:chart-consistency} fails.}
\label{fig:procedure}
\end{figure*}

Flattening is the step that causes the loss. Replace a model's concurrency by its serialisations, in the proportions a log would record them, and the result is another model with the same traces. Fitness, precision, alignment cost, entropic relevance and earth-mover cost are all computed from a log and a model's traces. Each takes the same value on both, so none rates the concurrent model above the serialised one. The directly-follows table, the object commercial tools compute first \citep{vanderaalstPractitionersGuideProcess2019}, is coarser still. On five activities each occurring once it has rank twenty where the stochastic language has rank one hundred and twenty, and a concurrency a miner reads off it comes from that miner's rule (Section~\ref{sec:dfg}). Concurrency in a discovered model is declared and not observed. Every other instrument entering an operational decision is quoted with a resolution, the smallest difference it can register, and here the chart alone fixes it.

Process mining names the model side of this effect representational bias \citep{aalstProcessMiningData2016}, the limit a discovery algorithm's chosen notation places on what it can express. The data side, the total order a log imposes on events that may have run in parallel, is treated separately under partially ordered event data \citep{leemansPartialorderbasedProcessMining2023}. Neither is given a form computable before a log is acquired. Representations keeping more than the traces fix their memory as an order chosen in advance, with no construction certifying that the chosen depth suffices. Named comparison distances are reported as single numbers, the representation each depends on left unstated. The within-execution concurrency needed to test any of this is scarce in public logs, so a ground truth has to be constructed rather than found.

We describe a process intrinsically as a distribution over labelled partial orders, each recording which work may run in parallel and which in sequence. Such a distribution carries both the concurrency a sequence cannot express and the frequencies that tradition leaves out. Every metric-shaped comparison of two such models makes two decisions that tools take jointly and report as one, a \emph{chart}, the concrete structure the model is rewritten in, and a \emph{geometry} on the structures the chart produces. On a simulated clinical fleet whose ground truth is fixed by construction, choosing the chart before the distance recovers all four generating groups at an adjusted Rand index of 1.000, where the best route through a flattened log stops at 0.633.

The set $\Nset$ of canonical labelled posets is read off an already mined model by the cospan presentation of \citet{leeStringDiagramsProcess2026}, and a model is a point of the simplex over $\Nset$ (Equation~\eqref{eq:model-vector}). The distance between two models is the pullback of the geometry $\mu$ along the chart $r$ (Equation~\eqref{eq:pullback}). The geometry is the Bhattacharyya angle of \citet{leeDistanceFunctionStochastic2026}, applied to language vectors directly and aggregated over rows for matrix-valued charts, and is taken as given.

Three consequences follow before any data is seen, so the choice of comparison can be stated in a study design and audited afterwards. The chart alone fixes which differences a comparison can register, and two models it sends to the same structure are two models no geometry separates, the models a chart conflates being its \emph{kernel}. The two pathways of Figure~\ref{fig:procedure}(a) are identified by every chart built from traces, and a chart that identifies the pathways a question is about cannot answer it, however large the log. Charts are ordered by how much they discard, so a difference registered on a coarse chart survives on every finer one, and only a difference that fails to register calls for a more detailed comparison. The pullback is a metric exactly when the chart discards nothing, and a pseudometric otherwise, so a pair of pathways reported as identical reflects a stated property of the chart rather than an accident of the data.

Figure~\ref{fig:procedure}(b) orders the work into four stages, whose named alternatives are catalogued in Section~\ref{sec:dictionary}, and the computational stages are implemented in the \texttt{proc-posets} package released with this paper.

The contributions of this paper are the following.

\begin{enumerate}
  \item \textbf{A concurrency supplied by the miner.} A directly-follows graph counts how often one activity follows another. Replace concurrent work by the orders a log would record it in and those counts do not move, so every graph a model produces is produced also by a model with no concurrency in it (Section~\ref{sec:dfg}). Theorem~\ref{thm:dfg-rank} measures how much the graph discards against what the traces keep. A concurrency a miner reports therefore comes from its own rule, and nothing in the traces confirms it. The timing the log already records carries the evidence a trace does not, and Section~\ref{sec:experiments} compares the two readings of one fleet.
  \item \textbf{The published measures, each at its own resolution.} Fitness, precision, entropic relevance and the earth-mover distance each rewrite a model before measuring it. A dictionary (Section~\ref{sec:dictionary}, tabulated in the supplementary material) names the rewriting each one performs and the differences that rewriting removes, and orders them by how much they discard (Theorem~\ref{thm:monotone}). A result already published stays valid inside the resolution the dictionary assigns it.
  \item \textbf{The limit a trace-level log imposes, in computable form.} Each variant leaving some order unconstrained contributes one relation, and those relations span the differences no trace-level comparison can register (Theorem~\ref{thm:kerL-linrel}). The span has a dimension and a membership test, so the resolution of a planned comparison can be written into a study design before any data is collected.
  \item \textbf{A representation that keeps everything, at a depth the model fixes.} A construction summarises a model by a tree over blocks of simultaneous work, and it stops on its own (Theorem~\ref{thm:faithful}). No memory length is fixed in advance and no parameter is tuned, so for a fixed comparison the search for a sufficient representation ends.
  \item \textbf{The staffing consequence, and an evaluation against a constructed ground truth.} Two steps run together and the same two steps run one after the other need the same total staff time, and differ in how much of it is needed at once. A capacity model reading the log only through total staff time is constant on the pair, so the conflation reaches the staffing decision (Section~\ref{sec:experiments}). On a simulated multi-site clinical fleet the partial-order chart recovers all four generating groups at an adjusted Rand index of 1.000, where the best route through a flattened log stops at 0.633. None of the public logs screened poses the comparison under a known generating model, so the ground truth is constructed.
\end{enumerate}

This paper introduces no discovery algorithm and no new distance. It states what no discovery algorithm can supply from sequences alone. It reports each comparison with its resolution, computed in advance of the data. Health care is our running setting, and the analysis applies to process comparison generally, in operational, business or industrial settings.

Section~\ref{sec:related-work} places the framework in the literature, Section~\ref{sec:model-vector} constructs the model vector and the two base charts, Section~\ref{sec:kernel-calculus} the kernel calculus and its geometry, Section~\ref{sec:dictionary} the dictionary, Section~\ref{sec:data} estimation from data, Section~\ref{sec:experiments} the clinical-fleet evaluation, and Section~\ref{sec:discussion} concludes. Proofs are collected in the supplementary material, together with the full dictionary table and the experimental apparatus.

\section{Related work}
\label{sec:related-work}

Operational research makes a statistical distance between data-generating distributions the primitive of a decision instrument \citep{Corne_2012, wangWassersteinDistributionallyRobust2020,ketkovStudyDistributionallyRobust2024,morganStatisticalProcessControl2025}. Of the two nearest works, one scores providers under a fixed Markov model instead of pathways \citep{topuzMarkovianScoreModel2024}, and the other compares cohorts on clinical logs but reports statistical significance instead of a distance \citep{mazharStochasticAwareComparativeProcess2023}. Neither addresses which representation and comparison to use, a choice that should follow the operational decision and not goodness of fit \citep{denboerDecisionbasedModelSelection2021}. A distance selected inside a tool is a modelling assumption entering an operational decision unrecorded, and two analysts may separate the same pair of sites or fail to with the disagreement visible in neither number. Sections~\ref{sec:kernel-calculus} and~\ref{sec:dictionary} make the choice explicit, by computing what each candidate comparison cannot separate. Selecting the method before applying it, instead of reporting the one that scored best afterwards, is a recognised contribution in that literature \citep{vanbulckWhichAlgorithmSelect2024}.

We organise the literature by position relative to $r$. A method may choose the pair $(r,\mu)$ and compute inside the pullback, may act upstream of $r$ by fixing what the observable is before any comparison applies to it, or may sit outside, returning a difference or a verdict with no $\mu$ pulled back. An upstream choice bounds every comparison downstream of it, and the object-centric and timestamp literatures therefore carry as much weight below as the distance literature does. Two adjacent lines fall outside our scope. Trace clustering groups the executions of one log \citep{evermannClusteringTracesUsing2016,feiClusteringPatientsTrajectories2013} rather than comparing models, though Section~\ref{sec:experiments} clusters sites once a distance between them exists. Measures of model syntax compare notation rather than behaviour, and are largely not metrics \citep{schoknechtSimilarityBusinessProcess2018}.

Process mining is dominated by discovery \citep{augustoAutomatedDiscoveryProcess2019} and conformance. Comparison within a common representation is less developed, usually cast as repository similarity search \citep{schoknechtSimilarityBusinessProcess2018}. The directly-follows graph is the object most tools compute first, and the practitioner literature documents its limitations as a catalogue of misleading cases \citep{vanderaalstPractitionersGuideProcess2019}, which is different in kind from a statement of what the graph cannot separate. Theorem~\ref{thm:dfg-rank} gives the latter. Discovery itself has been posed as an operational-research problem, continuously refitting a causal net as a non-stationary process drifts \citep{potoniecContinuousDiscoveryCausal2022}. Our partial-order approach follows a recent tradition in process mining \citep{leemansPartialorderbasedProcessMining2023}. Stochastically known logs \citep{galEverythingThereKnow2023} report an event attribute as a distribution over values rather than a single value. Section~\ref{sec:discussion} places that programme in this frame.

Upstream of $r$, object-centric event data \citep{bertiOCELObjectCentricEvent2024,vanderaalstObjectCentricProcessMining2023} extends the traditional event log with an oracle that does not consult the standard clock. Two events touching no common object impose no order on each other, so such a log presents an execution as a partial order (the channel-relative boundary, Section~\ref{sec:kerT}). Object-centric models, and traditional ones as a special case, are both describable natively as partial orders, a description the compositional framework of cospan algebras \citep{leeStringDiagramsProcess2026} recovers. Such processes are at present compared by slicing a log into sublogs placed side by side with no distance between them \citep{ghahfarokhiProcessComparisonUsing2021}. That literature is still settling which events constitute one execution. The choice, its \emph{case notion}, fixes what a variant partial order is \citep{adamsDefiningCasesVariants2022}. Section~\ref{sec:discussion} places the choice inside the representation, so competing definitions can be weighed on an equal footing.

Discovery methods for those models reach the object structure through a flattening. The Petri net construction of \citet{vanderaalstDiscoveringObjectCentricPetri2020} and the causal net construction of \citet{lissObjectCentricCausalNets2025} both flatten the log once per object type, mine each flat log with a conventional trace-level miner, and recombine the results, their reference implementations, which we read, flattening explicitly. The merge yields the relation between types, so two activities touching no common object are separated by it. Concurrency inside one type is settled earlier, by the per-type miner, from the frequencies with which two activities are seen in either order. That statistic is constant on $\ker\rT$, the kernel of the stochastic-language chart of Section~\ref{sec:stochastic-language-map}, taking the same value on a genuine concurrency and on the interleaving of Theorem~\ref{thm:kerL-linrel} matching its stochastic language, so the oracle the data supplies is exhausted before the concurrency question is put. The object-centric survey~\citep{bertiAdvancementsChallengesObjectCentric2023} has no statistical vocabulary, and the stochastic process mining survey~\citep{leemansTKDEsurvey} calls the object-centric extension open and underexplored, motivating the estimation results of Section~\ref{sec:data}.

Most of that tradition assigns no probability law to its linear extensions from which a metric could follow, though the field has its own identifiability theory, under which rediscoverability holds only up to trace equivalence \citep{leemansScalableProcessDiscovery2018}. The exception predates the field. \citet{mannila2000global} let a partial order generate its linear extensions uniformly and fit a mixture of such orders to sequence data by likelihood, searching series-parallel orders because counting linear extensions is $\#$P-complete in general, which is the model of Section~\ref{sec:model-vector} with the estimator of Section~\ref{sec:data}. Their score is a function of the observed sequences alone and is therefore constant on the fibres of $\rT$, and their fitted family carries one non-trivial order against a uniform background, so the mixture of chains matching that order lies outside the family. Identifiability is not raised there, and Section~\ref{sec:kernel-calculus} supplies it. The closest description to our own is \emph{possibilistic}, a canonical normal form rewriting interleaving into concurrency at atomic granularity \citep{leemansInformationPreservingAbstractions2020}. We strengthen this to include observed trace frequencies, with a kernel basis and a rank (Section~\ref{sec:kernel-calculus}).

The nearest stochastic statement is partially ordered stochastic conformance checking \citep{leemansPartiallyOrderedStochastic2025}. Its motivating observation is that a partial order distributes probability across its linear extensions, so the interleaving carries no information beyond the order (their Problem P2 and Lemma 2). That redundancy is what Theorem~\ref{thm:kerL-linrel} rests on. We use it for a negative result, that genuine concurrency and a balanced choice of the same steps' orders induce the same stochastic language.

Inside the pullback, the stochastic and behavioural literature largely chooses $(r,\mu)$ on the causal-poset object, recovered with its kernel in Section~\ref{sec:dictionary} \citep{liJensenShannonDistance2025,KemenySnell1962,weidlichEfficientConsistencyMeasurement2011,kunzeBehavioralSimilarityProper2011,leemansStochasticProcessMining2021}. Outside it, comparison on canonically reduced event structures \citep{armas-cervantesBehavioralComparisonProcess2014} returns an interpretable difference rather than a magnitude, and is the closest of these to our keeping concurrency atomic. Comparisons returning only the verdict of a test are placed in Section~\ref{sec:discussion}.

Markovian abstractions of order $k$ underlie abstract-and-compare precision measures, comparing model and log by multisets of length-$k$ subtraces \citep{augustoAbstractandCompareFamilyScalable2018,augustoMeasuringFitnessPrecision2022}. A variable-length Markov chain fitted to a log \citep{incertoStochasticConformanceChecking2025} selects its contexts from data by a statistical criterion, using the observed activities only. Three improvements are needed to describe more realistic processes. States should be canonical modular blocks kept atomic, so concurrency is never serialised, the summarisation scheme being the one those methods already use and the alphabet being what changes. Identification should be injective at a computed memory depth, not a fixed order. The representation should account for smooth changes in probabilities, not a set-difference count. Section~\ref{sec:model-vector} supplies the first, Theorem~\ref{thm:faithful} the second, and the geometry of Section~\ref{sec:geometry} the third.

\section{The model vector and our two base charts}
\label{sec:model-vector}

This section fixes the model vector and the two base charts of its title, the modular chart and the stochastic-language map. The input to this paper is a process model already mined and converted to a poset representation by the cospan approach of~\citet{leeStringDiagramsProcess2026}, which presents Petri nets, causal nets, process trees and BPMN diagrams alike. A variant $D$ is a labelled poset on its set of occurrences $\actset_D$, over the activity alphabet $\alphabet$. Its linear extensions $\lin{D}$ are the total orders of those occurrences respecting the causal order, and the labelling $\ell\colon\actset_D\to\alphabet$ turns such an order into a trace, a word over $\alphabet$. A trace is written in angle brackets, $\trace{acba}$, in which the label $a$ repeats.

\subsection{Variants as labelled posets}

\begin{definition}[Intrinsic object of a variant]
\label{def:intrinsic}
The \emph{intrinsic object} of a variant $D$ is its labelled poset up to isomorphism.
\end{definition}

A partial order records which steps of a single execution must occur in order, leaving the rest free. The poset is the compared object, so the firing sequences come from $D$ alone, and a partial order is determined by its linear extensions~\citep{szpilrajnExtensionOrdrePartiel1930,gischerEquationalTheoryPomsets1988}. A distribution over $\lin{D}$ is written $\nu_{D}$. The letter $D$ denotes a finite labelled causal poset throughout, and the domain fixes which posets are meant, $D\in\Nset$ where membership of the model space is required and an arbitrary $D$ otherwise.

\subsection{The model vector}

Let $\Nset$ be the set of these canonical posets, one per isomorphism class, and write $e_{D}$ for the basis vector indexed by the variant $D$. We write $\rho$ for a distribution over a finite candidate set of variants, presenting a model as a point of the simplex $\simplexN$ inside the free real vector space $\modelspace$,
\begin{equation}
\label{eq:model-vector}
  m\;=\;\sum_{D\in\Nset}\rho(D)\,e_{D}\;\in\;\simplexN\subset\modelspace ,
\end{equation}
with $\rho(D)\ge0$ and $\sum_{D}\rho(D)=1$. The vectors $\{e_{D}\}_{D\in\Nset}$ form a basis because distinct posets are distinct atoms (Definition~\ref{def:intrinsic}), and each $e_{D}$ is a vertex of $\simplexN$, so no linear relation among them is imposed. Every construction below uses a model only through Equation~\eqref{eq:model-vector}, never a syntactic presentation.

\subsection{Iteration and silent steps}
\label{sec:iteration}

A looping model is unrolled by indexing the loop by its repeat count $n\in\mathbb{N}$. Each $n$ gives one acyclic variant $D_{n}$, the loop body composed $n$ times in sequence (the signature, which records a model's activities and the typed interfaces along which they compose~\citep{leeStringDiagramsProcess2026}, realises this closure as self-composition of the loop-entry generator), and the model is the mixture over depths, a formal sum until the cut-off below makes it finite, \[ m\;=\;\sum_{n\in\mathbb{N}}\rho(D_{n})\,e_{D_{n}} . \] The weight $\rho(D_{n})$ is nonzero for only finitely many $n$, so $\Nset$ stays finite and every construction below applies to the unrolled family verbatim. With an event log, no trace repeats the body beyond some observed maximum, so $\rho$ is supported on those depths. Without a log, a generative assumption gives a geometric law $\rho(D_{n})\propto(1-p)^{n}$, truncated where the residual tail mass falls below a stated tolerance. Silent (invisible) steps, unlabelled transitions carrying no activity, do not arise in the compared object, since the signature extraction of~\citet{leeStringDiagramsProcess2026} removes them before this paper begins.

\subsection{The modular chart}
\label{sec:tiling}

A universal map gives posets coordinates in which variants sharing a sub-pattern share a coordinate, decomposing a poset canonically into pieces run in sequence (series), pieces run in any order (parallel), and a residual ``prime'' piece for any fragment that is neither, the smallest example being the classic $N$-poset. The first two are compositions of posets, in which the decomposition below and the examples throughout are written.

\begin{definition}[Series and parallel composition]
\label{def:series-parallel}
For finite labelled posets $D$ and $D'$ on disjoint sets of occurrences, the \emph{series composition} $D;D'$ is their union with every occurrence of $D$ ordered before every occurrence of $D'$, and the \emph{parallel composition} $D\otimes D'$ is their union with no order between the two sides. Both keep the internal orders and the labels~\citep{gischerEquationalTheoryPomsets1988}. A bare activity $a$ denotes the one-occurrence poset labelled $a$, so $a;b;c$ is the chain spelling $\trace{abc}$ and $a;(b\otimes c)$ orders $a$ before $b$ and $c$ while leaving $b$ and $c$ incomparable.
\end{definition}

\begin{proposition}[Gallai tiling~\citep{gallai_transitiv_1967}]
\label{prop:canonical-tiling}
Every nonempty finite labelled poset has a unique reduced modular-decomposition tree whose internal nodes are series, parallel, or prime. Collecting the blocks observed across the model class $\Nset$ (the singleton leaves and the maximal parallel or prime nodes, each taken up to isomorphism) into an alphabet $\blockset=\{\beta_{1},\beta_{2},\dots\}$, the root of the tree gives a block word $\tile(D)=\beta_{i_{1}};\cdots;\beta_{i_{m}}$, and $\tile$ is injective on $\Nset$.
\end{proposition}

Both operations of Definition~\ref{def:series-parallel} are associative and $\otimes$ is commutative, so the block word $\tile(D)$ denotes the series composition of its blocks and concatenation of block words is again series composition. The posets generated from bare activities by the two operations are exactly the series-parallel ones, those whose modular-decomposition tree has no prime node~\citep{valdesRecognitionSeriesParallel1982,gischerEquationalTheoryPomsets1988}. A prime module is an irreducible object of the decomposition, entering the block alphabet $\blockset$ as a single letter.

\begin{definition}[Modular chart]
\label{def:modular-chart}
The chart $\rB$ sends each variant to its block word and rewrites the model in that basis, so that
\begin{equation}
\label{eq:rB}
\begin{aligned}
  \rB(e_{D})\;&=\;e_{\tile(D)} , \\
  \rB(m)\;&=\;\sum_{D\in\Nset}\rho(D)\,e_{\tile(D)} .
\end{aligned}
\end{equation}
\end{definition}

Since $\tile(\cdot)$ is injective on $\Nset$, $\rB$ is a linear injection, hence distance-preserving for any coordinate-wise metric. The block word is therefore the variant in other coordinates, and a distribution over posets is a distribution over words in the block alphabet. The block sequence depends on the order alone, so the chart follows any relabelling of activities.

\subsection{The stochastic-language map}
\label{sec:stochastic-language-map}

A variant's stochastic language needs a distribution over its linear extensions, a scheduling law. The model space carries one such law, fixed by Assumption~\ref{ass:fixed-law}, so $\rT$ takes the model alone. A statement needing the uniform law cites Assumption~\ref{ass:uniform}.

\begin{definition}[Scheduling law]
\label{def:scheduling-law}
A scheduling law assigns to each finite causal poset $D$ a distribution $\nu_D\in\Delta(\lin{D})$ over its linear extensions, depending on $D$ only through its isomorphism class.
\end{definition}

The law $\nu_{D}$ is not itself a distribution over traces, since a label function may collapse several linear extensions to one fibre. Every construction below holds for any fixed scheduling law, a concrete choice being needed only for numerical values and for the log-side estimation. When needed, we work with two modelling assumptions, the first fixing when the law is chosen and the second which law it is. Choosing it in advance keeps the weight likelihood of Section~\ref{sec:data} a mixture with known components, and so concave, and settles the trace-level boundary of Section~\ref{sec:kerT} before the log is opened.

\begin{assumption}[Fixed scheduling law]
\label{ass:fixed-law}
The scheduling law is fixed before the data and not estimated alongside the weights.
\end{assumption}

\begin{assumption}[Uniform scheduling law]
\label{ass:uniform}
We take the scheduling law to be uniform. Conditional on $D$, every $\sigma\in\lin{D}$ has equal probability $\nu_{D}(\sigma)=1/\lvert\lin{D}\rvert$.
\end{assumption}

The uniform law is the model-side counterpart of the uniform interleaving partial-order resolution places over order-ambiguous events in a log~\citep{vanderaaPartialOrderResolution2020}, and it invokes the principle of indifference in its maximum-entropy form~\citep{jaynesInformationTheoryStatistical1957,coverElementsInformationTheory2006}, posets carrying ordering information alone. The energy tilt $\nu_{D}^{\vartheta}(\sigma)\propto e^{-\vartheta E(D,\sigma)}$, for a fixed energy $E(D,\sigma)\in\mathbb{R}$ scoring each linear extension $\sigma\in\lin{D}$, is a standard alternative, recovering the uniform law at $\vartheta=0$. The uniform law over the linear extensions of a mixture of series-parallel posets is also the generative model \citet{mannila2000global} fit to sequence data, where the weights are estimated by likelihood as in Section~\ref{sec:data}.

An \emph{end-to-end execution} of the process $m$, an \emph{execution} for short, is a draw of a variant $D$ with chance $\rho(D)$. The occurrences of $D$ are distinct, so a linear extension $\sigma\in\lin{D}$ is a permutation of them without repeats, the execution's total order, and the \emph{trace} of the execution is the image $\ell(\sigma)$ under the labelling. A log is a sample of traces, summarised by its empirical distribution. A trace may repeat a label, so several extensions can spell one, and extending $\ell$ letterwise the fibre $\ell^{-1}(w;D)=\{\sigma\in\lin{D}:\ell(\sigma)=w\}$ collects them, the mass the scheduling law places on it setting the probability of $w$.

\begin{definition}[Stochastic-language map]
\label{def:stochastic-language-map}
\sloppy
The \emph{stochastic-language map} $\rT\colon\modelspace\to\mathbb{R}^{\alphabet^{*}}$ is the linear map fixed on the variant basis by the pushforward of the scheduling law along $\ell$, the \emph{variant law} $\rho(\cdot\mid D)=\ell_{*}\nu_{D}$. Writing $e_{w}$ for the point mass on the trace $w$, its coefficient is the mass the scheduling law places on the fibre over $w$,
\begin{equation}
\label{eq:variant-law}
\begin{aligned}
  \rT(e_{D})\;&=\;\sum_{w\in\alphabet^{*}}\rho(w\mid D)\,e_{w} , \\
  \rho(w\mid D)\;&=\;\nu_{D}(\ell^{-1}(w;D))\;=\!\!\!
    \sum_{\substack{\sigma\in\lin{D}\\ \ell(\sigma)=w}}\!\!\!\nu_{D}(\sigma) .
\end{aligned}
\end{equation}
Extending linearly over Equation~\eqref{eq:model-vector} gives the \emph{stochastic language} of a model $m$, its coefficients in the trace basis,
\begin{equation}
\label{eq:marginal}
\begin{aligned}
  \rT(m)\;&=\;\sum_{w\in\alphabet^{*}}\rho(w)\,e_{w} , \\
  \rho(w)\;&=\;\sum_{D\in\Nset}\rho(D)\,\rho(w\mid D) .
\end{aligned}
\end{equation}
\end{definition}
Equation~\eqref{eq:model-vector} expresses a model in the variant basis and Equation~\eqref{eq:marginal} its language in the trace basis, where $e_{w}$ is the image under $\rT$ of the chain variant spelling $w$, the unique variant with a single linear extension. The schedule $\nu$ supplies probability, while $\ell_{*}$ is deterministic and only collapses the extensions that spell the same word. Under Assumption~\ref{ass:uniform} every linear extension in a fibre has the same weight, so the variant law becomes a ratio of counts,
\begin{equation}
\label{eq:uniform-law}
  \rho(w\mid D)\;=\;\frac{\lvert\ell^{-1}(w;D)\rvert}{\lvert\lin{D}\rvert} ,
\end{equation}
and the stochastic language mixes these with the variant weights $\rho(D)$.

Unlike $\rB$, $\rT$ takes the scheduling law as an input and is not injective, so distinct models can share a stochastic language, worked out in full as Example~\ref{ex:block-matrix}. Section~\ref{sec:kerT} characterises those coincidences in general.

\section{The kernel calculus}
\label{sec:kernel-calculus}

Section~\ref{sec:intro} described a comparison as a chart and a geometry, the pair $(r,\mu)$, in which a chart $r$ rewrites the model in terms of a concrete structure, a fixed metric $\mu$ measures the images, and the distance between two models is the pullback of $\mu$ along $r$,
\begin{equation}
\label{eq:pullback}
  d(m,m')\;=\;\mu\bigl(r(m),\,r(m')\bigr).
\end{equation}
The pullback is written with the chart's subscript, so $\dT(m,m')=\dBA\bigl(\rT(m),\rT(m')\bigr)$, and, for the block chart of Definition~\ref{def:block-matrix}, $\dP(m,m')=\dSMD\bigl(\rP(m),\rP(m')\bigr)$.

For a metric $\mu$ the pullback~\eqref{eq:pullback} is a pseudometric vanishing on the fibres of $r$, and is a metric if and only if $r$ is injective~\citep{buragoCourseMetricGeometry2001,dezaEncyclopediaDistances2016}. The fibres fix which models the comparison distinguishes, and ``kernel'' names that fibre relation, the linear-subspace definition being reserved for genuinely linear charts. For $\rT$, linear on the variant basis, $\ker\rT=\{0\}$ iff the variant laws $\{\rho(\cdot\mid D)\}_{D\in\Nset}$ of Definition~\ref{def:stochastic-language-map} are linearly independent, a condition on the image alone. This section fixes the default geometry, computes the kernel of each chart of Section~\ref{sec:model-vector}, bounds every comparison under a stochastic map, and closes by carrying the whole construction to object-centric logs (Section~\ref{sec:object-centric}).

\subsection{Fisher--Rao as the default geometry}
\label{sec:geometry}

By default we take for $\mu$ the Fisher--Rao geodesic distance on the categorical simplex~\citep{amariInformationGeometry2021}. Write $\bc(p,q)=\sum_{x}\sqrt{p(x)\,q(x)}$ for the Bhattacharyya coefficient of two probability vectors on a common set. The angle between them is
\begin{equation}
\label{eq:dBA}
  \dBA(p,q)\;=\;2\arccos\bc(p,q).
\end{equation}
On two block-transition matrices $\smat,\smat'$ over a common state space $X$ we use the framework of~\citet{leeDistanceFunctionStochastic2026}, applying it row by row in the normalised version
\begin{equation}
\label{eq:dSMD}
  \dSMD(\smat,\smat')\;=\;2\sqrt{\frac{1}{\lvert X\rvert}\sum_{i\in X}\arccos^{2}\!\Big(\sum_{j}\sqrt{\smat_{ij}\,\smat'_{ij}}\Big)},
\end{equation}
the root-mean-square row angle. The $1/\sqrt{\lvert X\rvert}$ factor renders distances comparable across state-space sizes. The \emph{comparison state space} $X$, the union of the block-chart states over every model compared, is fixed once and never rebuilt per pair, so $\dSMD$ is a fixed rescaling of the row-wise Fisher--Rao metric and itself a metric.

Both distances share a scale, $[0,\pi]$, zero at agreement and $\pi$ at disjoint support. On a finite sample space the Fisher--Rao metric is, up to scale, the unique Riemannian metric contracting under every stochastic map (Chentsov's theorem, \citealp{cencovStatisticalDecisionRules1982,amariInformationGeometryIts2016}). That uniqueness is relative to its Riemannian and monotonicity hypotheses. Separately, the metric bounds the depth-truncation error of Section~\ref{sec:iteration} at $O(\sqrt{\varepsilon})$ in the discarded mass $\varepsilon$, a bound the triangle inequality extends to every distance computed from the truncated model. The choice is a default. Section~\ref{sec:dictionary} catalogues named comparisons that hold a chart fixed and vary $\mu$.

Equation~\eqref{eq:dBA} compares a log and a model's stochastic language in the trace simplex. Equation~\eqref{eq:dSMD} is the default model--model distance, on block-transition matrices, and applies once a model has been discovered. Computing~\eqref{eq:dBA} for a general poset is $\#$P-complete~\citep{brightwellCountingLinearExtensions1991a}, a cost any process model meets and not peculiar to this representation. The block chart does not inherit it, blocks staying atomic, so $\rP$ is linear-time on a prime-free tiling (Section~\ref{sec:tiling}).

\subsection{The kernel of the stochastic-language chart}
\label{sec:kerT}

Which models $\rT$ identifies is fixed, at the linear level, by one family of relations. A chain poset variant spells a single word, $\rT(e_{c_{w}})=e_{w}$, and the relations below express a concurrency through the chains spelling its traces, so those chains must be available to write them with.

\begin{definition}[Chain-complete model space]
\label{def:chain-complete}
A model space $\Nset$ is \emph{chain-complete} when it contains the chain $c_{w}$ of every word $w$ any of its member posets can emit. A model space is completed by adjoining the missing chains.
\end{definition}

Chain-completeness may be assumed without loss of generality. The adjoined chains carry no weight, so every model and language is unchanged, and they biject with the word basis of $\operatorname{im}\rT$ used below. Completion enlarges $\ker\rT$ without creating a coincidence and changes neither $X$ nor the scale of~\eqref{eq:dSMD}, and the same device covers a trace observed but not previously modelled.

\begin{lemma}[The chain--kernel splitting]
\label{lem:chain-kernel-split}
Let $\Nset$ be chain-complete, let $\mathcal{C}\subseteq\mathbb{R}^{\Nset}$ be the span of the chain atoms, and write $\seqmap(m)=\sum_{w}\rT(m)(w)\,e_{c_{w}}$ for the mixture of chains spelling the stochastic language of $m$. Then $\rT$ restricts to a bijection of $\mathcal{C}$ onto $\operatorname{im}\rT$, the vector space splits as $\mathbb{R}^{\Nset}=\mathcal{C}\oplus\ker\rT$ with $\operatorname{rank}\rT=\#\{\text{chain atoms}\}$ and $\dim\ker\rT=\lvert\Nset\rvert-\#\{\text{chain atoms}\}$, and $\seqmap$ is the projection onto $\mathcal{C}$ along $\ker\rT$.
\end{lemma}

The chains therefore carry the whole of what a trace-level log records, and $\seqmap(m)$ is the concurrency-free model reproducing the language of $m$. No such log registers the difference between a variant and its own reconstruction, and the theorem below states that those differences form a basis of $\ker\rT$.

\begin{theorem}[The linearisation relations span \texorpdfstring{$\ker\rT$}{ker r\_T}]
\label{thm:kerL-linrel}
Assume a model space is chain-complete. The linearisation relations, one for each non-chain $D\in\Nset$,
\begin{equation}
\label{eq:linrel}
  \linrel[D]\;=\;e_{D}\;-\;\sum_{w}\rho(w\mid D)\,e_{c_{w}},
\end{equation}
form a basis of $\ker\rT$.
\end{theorem}

By Theorem~\ref{thm:kerL-linrel}, two mixtures of poset variants (Equation~\eqref{eq:model-vector}) carry the same stochastic language exactly when they differ by an element of the span of the linearisation relations, the identifiability limit of the trace-level pullback.

\begin{example}[The concurrency--interleaving family]
\label{ex:block-matrix}
Consider three variants $a;(b\otimes c)$, $a;b;c$ and $a;c;b$ together with weights $\rho=(p+q,\tfrac12-p,\tfrac12-q)$, which run over the whole simplex as $p$ and $q$ range over $p\le\tfrac12$, $q\le\tfrac12$ and $p+q\ge0$, and let the scheduling law place $\theta$ on $\trace{abc}$ inside the block. Writing $s=(1-\theta)p-\theta q$, the stochastic language is $(\tfrac12-s)\,e_{abc}+(\tfrac12+s)\,e_{acb}$, so the weights reach the log through $s$ alone and the language is constant along the ridge $(p,q)\propto(\theta,1-\theta)$. The concurrency $a;(b\otimes c)$ sits at $p=q=\tfrac12$, the balanced interleaving $\tfrac12\,a;b;c+\tfrac12\,a;c;b$ at $p=q=0$, and under the uniform law the ridge joins them.
\end{example}

Under the uniform law, a genuine concurrency and a balanced choice between its two orders are therefore one object to any comparison built from traces, separated by the single linearisation relation $\linrel[a;(b\otimes c)]=e_{a;(b\otimes c)}-\tfrac12(e_{a;b;c}+e_{a;c;b})$ and by nothing a log can record.

A flat log does not decide between a concurrency and an interleaved choice, so a method reporting one has answered from an assumption of its own~\citep{aalstProcessMiningData2016}. A log's empirical stochastic language is always realised by some concurrency-free model, the chain mixture of Lemma~\ref{lem:chain-kernel-split}, and every score computed from a log and a model's stochastic language alone is constant across that fibre. None can therefore rate a discovered concurrency above that concurrency-free model, and the same holds for directly-follows discovery (Section~\ref{sec:dfg}). The supplementary material measures one such displacement on the trace-identical arm of Section~\ref{sec:exp-graceful}, where a miner's induced model agrees with the generating model on this fibre at every mixing weight and the entire error is confined to it.

\begin{corollary}[The trace-level boundary]
\label{cor:lin-boundary}
Call a comparison trace-observable if $r=g\circ\rT$ for some map $g$. On a chain-complete model space (Definition~\ref{def:chain-complete}), whenever some variant is non-chain, no trace-observable comparison is a metric, since none separates a genuine concurrency from the interleaving matching its stochastic language. A representation lies on the boundary when $\ker r=\ker\rT$, and above it when $\ker r\subsetneq\ker\rT$. Both $\rT$ and the earth-mover comparison under any ground cost separating distinct traces attain the boundary.
\end{corollary}

The boundary bounds $\rT$ alone. A comparison built on any channel separates nothing in that channel's kernel, so a finer channel lowers the boundary, and object references keep the clock's order only between events sharing an object. The supplementary material records what this means for timestamps, for models fitted to flat traces, and for the scheduling law. The classification is not exhaustive. A chart may separate models within a fibre of $\rT$ while conflating models from different fibres, and is then off the boundary, as the activity footprint of Section~\ref{sec:dictionary-rows} is.

\subsection{The block chart}
\label{sec:block-chart}

By Corollary~\ref{cor:lin-boundary}, a decision depending on a direction of $\ker\rT$ needs a chart above the boundary. The modular chart $\rB$ is injective and so lies above it, but represents a model as a distribution over whole block words, so two models meet only as whole executions. The chart of this subsection is injective and compares two models one state at a time. Its state space carries the concurrency that a Markov chain over activities discards.

\begin{definition}[Block chart]
\label{def:block-matrix}
The chart reduces a model to a Markov chain whose states are contexts of blocks. The distribution $\rB(m)$ over block words is summarised by the context-tree construction of \citet{buhlmannVariableLengthMarkov1999}, applied to the exact weights of the model vector. No selection criterion enters and the summary is a function of $\rB(m)$ alone.

A \emph{history} is a run of blocks that some variant's word begins with, and its \emph{continuation law} gives the conditional weight of each further run and of stopping. A \emph{context} is a run of blocks, \emph{safe} when every history ending in it carries the same continuation law. The context tree $\ctree$ assigns to each history the shortest safe context ending it, and the history itself where none is safe. Every context is nonempty, so it ends in the block just emitted, and the empty history alone maps to the source $\gs$. The states are the contexts so reached together with $\gs$ and a stopping state $\gt$, and $\smat[u,\cdot]$ row-normalises the variant weight crossing each transition out of $u$.

Writing $\Phi$ for that summary, $\Phi(\rB(m))=\smat$ and $\rP=\Phi\circ\rB$. Limiting a context to $k$ blocks, so that a history with no safe context within $k$ takes its last $k$ blocks instead, gives $\Phik$ and the chart $\rP^{(k)}=\Phik\circ\rB$. The supplementary material carries the construction out on the two ends of Example~\ref{ex:block-matrix}, with the concurrency's matrix and the interleaving's unsafe contexts.
\end{definition}

The memory limit $k$ bounds how far back a state may look. Contexts below it keep their own differing lengths, so the states of a chart arrive at varied depths under any $k$, and the limit takes effect only where no safe context fits. Safety constrains the whole continuation, beyond the next block, since merging contexts that agree only on the next block can assign weight to block words $m$ never produces. Two conventions make $\smat$ row-stochastic on a shared state space (a return edge $\gt\to\gs$, a jump to $\gt$ from a state a model does not visit), fixed once per comparison from a context tree over every model compared (Proposition~\ref{prop:fleet-tree}). Blocks are atomic, never expanded into their interior interleavings, so the within-block uniform law (Assumption~\ref{ass:uniform}) plays no role, and the block chart is scheduling-law-free.

\begin{sloppypar}
Row-normalisation is a ratio, so $\rP$ is non-linear and its injectivity is a point-separation statement, not a rank count (Theorem~\ref{thm:faithful}). Both steps of $\rP=\Phi\circ\rB$ are lossless, $\rB$ as a change of basis and $\Phi$ because its merges are exactly those the continuation laws license. Detail is lost only under the truncation $\Phik$ with $k<\kstar$.
\end{sloppypar}

\begin{theorem}[Faithfulness]
\label{thm:faithful}
Write $L$ for the greatest number of blocks in the block word of a variant of $m$. The chart $\rP$ of Definition~\ref{def:block-matrix} is faithful, in that $m\mapsto\smat$ is injective on the models of a fixed signature. Under truncation to a memory limit the same holds at every $k\ge\kstar(m)$, where the least faithful limit satisfies $\kstar(m)\le L$ and is obtained by a finite search.
\end{theorem}

Injectivity is per-signature. A label-merging morphism need not preserve distinctness, and the chart commutes with one only when the merged rows already share an outgoing distribution.

\begin{proposition}[One tree for a comparison]
\label{prop:fleet-tree}
Let $m_{1},\dots,m_{M}$ be the models of a comparison and let $\ctree$ merge a context only where it is safe in every one of them. Then $\ctree$ is lossless for each $m_{i}$ separately, so Theorem~\ref{thm:faithful} holds on the common state space $X$ it defines, and $\kstar=\max_{i}\kstar(m_{i})$.
\end{proposition}

Building the tree over the whole comparison keeps $\lvert X\rvert$ one constant for the study. Enlarging $X$ beyond the states some member visits leaves the distinctions intact but changes the $1/\sqrt{\lvert X\rvert}$ scale, so faithfulness is relative to the comparison. A model admitted later can invalidate a merge, forcing a rebuild that may raise $\kstar$ and rescale every distance, and the class-wide bound $\kstar(m)\le L$ is therefore kept alongside.

\subsection{The kernel of the directly-follows map}
\label{sec:dfg}

The directly-follows graph (DFG) records, for each ordered activity pair $(x,y)$, how often $y$ directly follows $x$ in a flat log. It is a map of the model vector, the stochastic-language chart followed by a linear count of adjacent pairs, $\dfgmap=g\circ\rT$. Its target is a table of adjacency rates and not a law, so no $\mu$ of Section~\ref{sec:geometry} applies directly and it is placed here by its kernel. The table is the sole input of the directly-follows Inductive Miner~\citep{leemansScalableProcessDiscovery2018}. The Heuristics Miner's dependency counts~\citep{weijtersFlexibleHeuristicsMiner2011} are computed from one such table and nothing else. The Probabilistic Inductive Miner~\citep{bronsStrikingNewBalance2021} recomputes one per subtree, so the results below bound it one table at a time. It is a different object from the block-successor footprint, $\rP$ at $k=1$ (Section~\ref{sec:dictionary}), which records succession between \emph{blocks} of the tiling after the log is lifted to its partial order and cannot be recovered from a DFG.

\begin{definition}[The directly-follows map]
\label{def:dfg-map}
Let $D$ be a variant and $\widehat{D}$ its augmentation by a bottom and a top labelled outside $\alphabet$, so start and end activities become ordinary adjacencies. For distinct elements $u,v$ let $E_{uv}\subseteq\lin{\widehat{D}}$ be the linear extensions placing $v$ immediately after $u$, and put
\begin{equation}
\delta_{D}(x,y)=\sum_{u:\ell(u)=x}\ \sum_{\substack{v:\ell(v)=y\\v\neq u}}
\nu_{\widehat{D}}(E_{uv}),
\label{eq:dfg-entry}
\end{equation}
the weight the scheduling law $\nu$ places on those extensions, extended linearly by $\dfgmap(m)=\sum_{D}\rho(D)\,\delta_{D}$ over the model vector~\eqref{eq:model-vector}.
\end{definition}

Under the uniform law the weight is a ratio of linear-extension counts, obtained by contracting the pair $uv$. The supplementary material gives the formula and interprets $\delta_D(x,y)$, under every scheduling law, as the expected number of adjacent $xy$ occurrences in a $\nu$-random linear extension of $D$.

On a chain-complete model space (Definition~\ref{def:chain-complete}) the table is a function of the stochastic language, $\dfgmap=g\circ\rT$ with $g$ linear, so $\dfgmap$ is trace-observable in the sense of Corollary~\ref{cor:lin-boundary} and $\ker\rT\subseteq\ker\dfgmap\subseteq\ker(h\circ\dfgmap)$ for any further map $h$ of the table, a frequency threshold included. For the chain mixture $\seqmap$ of Lemma~\ref{lem:chain-kernel-split}, $\dfgmap(\seqmap(m))=\dfgmap(m)$. On any single shape (Theorem~\ref{thm:dfg-rank}), and under no hypothesis on labels, every table a model induces is induced also by a model carrying no concurrency. A concurrency reported by a directly-follows-based miner therefore comes from the miner's own rule, the $\alpha$-algorithm's parallel relation or the Inductive Miner's cut rules~\citep{aalstProcessMiningData2016}, and no function of the table certifies it.

\begin{theorem}[Rank and kernel of the directly-follows map]
\label{thm:dfg-rank}
Call the multiset of activity labels a variant carries its \emph{shape} $s$, $s(x)$ the number of occurrences of $x$, $\lvert s\rvert=\sum_x s(x)$, $n$ the number of activities occurring at all, and $s$ \emph{simple} when every $s(x)=1$. Assume $\lvert s\rvert\ge2$. Write $\Nset_s$ for the posets of shape $s$, whose chains biject with the distinct words $s$ spells. On $\mathbb{R}^{\Nset_s}$,
\begin{equation}
\begin{aligned}
\operatorname{rank}\rT&=\frac{\lvert s\rvert!}{\prod_{x:s(x)\ge1}s(x)!},\\
\operatorname{rank}\dfgmap&=n(n-1)+\#\{x:s(x)\ge2\},
\end{aligned}
\label{eq:dfg-ranks}
\end{equation}
so for a simple shape the two ranks are $n!$ and $n(n-1)$. The ranks are fixed by the chains, which carry weight $1$ under every scheduling law, so the statement holds on a single shape and for each $\nu$.
\end{theorem}

On the same shape the two losses compose as a direct sum, $\ker\dfgmap=\ker\rT\oplus(\ker\dfgmap\cap\mathcal{C})$, the linearisation relations of Theorem~\ref{thm:kerL-linrel} forming a basis of the first summand. The second summand vanishes on a shape of a single activity and on simple shapes of up to three activities and grows factorially thereafter, the gap between trace directions and distinguished ones widening with every activity added. Repeated activities enlarge the kernel rather than escape it, $e_{b;(a\otimes a);c}-e_{b;a;a;c}$ being a linearisation relation the graph inherits. A model spanning several shapes is a mixture across which $\dfgmap$ is linear, so~\eqref{eq:dfg-ranks} is applied shape by shape. This is the population form of process discovery's representational bias~\citep{aalstProcessMiningData2016}.

\subsection{Monotonicity under stochastic maps}
\label{sec:monotone}

A log is a sample from a law over traces. Any rule $\Lambda$ applied to one execution at a time and fixed in advance is a stochastic map, linear on laws, whether it merges activities, drops a resource label, buckets a duration, deletes an event or coarsens a timestamp. Theorem~\ref{thm:monotone} bounds all of them at once, no such rule increasing the Bhattacharyya angle between two laws. A rule fitted to the logs being compared is not covered. Fitting makes the rule a function of the data, so one execution's output depends on the rest of the sample, the map is no longer linear on laws, and the bound does not apply.

\begin{theorem}[Monotonicity under stochastic maps]
\label{thm:monotone}
Let $\Lambda$ be a stochastic map, a randomised post-processing that, given $x$, emits $y$ with probability $\Lambda(y\mid x)$, so $\sum_y\Lambda(y\mid x)=1$ for every $x$. For any laws $p,q$ on a countable set and any such $\Lambda$,
\begin{equation}
\label{eq:dpi}
  \dBA(\Lambda p,\Lambda q)\;\le\;\dBA(p,q),
\end{equation}
with $\dBA=2\arccos\bc$ the Bhattacharyya angle of Equation~\eqref{eq:dBA}.
\end{theorem}

Consequently, no transformation of the data chosen in advance can create a difference between two models, only shrink one, and the same holds row by row of $\dSMD$. A nonzero distance on one chart therefore remains nonzero on every chart from which that chart factors through such a map. Raising the memory limit is not a relation of that kind, so the conclusion does not extend along the memory axis. Only when an expected difference does not register should a more detailed chart be used.

\subsection{Object-centric event data on the same charts}
\label{sec:object-centric}

An object-centric log records which objects each event touches, and two events sharing no object are left unordered, so such a log already presents an execution as a partial order. The typed information enters as a refinement of the labels, so the constructions listed in Corollary~\ref{cor:object-centric} apply to it unchanged. One hypothesis is untouched by the enrichment, the independence of executions assumed in Section~\ref{sec:data}, which the execution notion, the case notion of the literature, decides.

\begin{corollary}[Object-centric variants are labelled posets]
\label{cor:object-centric}
Let each label of a variant be enriched to record the object types its event touches and the typed wires it closes, and let $\Nset$ be the resulting set of labelled posets. Proposition~\ref{prop:canonical-tiling}, Definition~\ref{def:modular-chart}, Theorem~\ref{thm:faithful} and Proposition~\ref{prop:fleet-tree} hold verbatim over the enriched labels, as do the reductions of Section~\ref{sec:dictionary} and Theorem~\ref{thm:monotone}. The enriched labels enlarge $\blockset$ and with it $\lvert X\rvert$, and $\kstar$ is recomputed on the enriched tree.
\end{corollary}

\section{A dictionary of named comparisons}
\label{sec:dictionary}
\label{sec:dictionary-rows}

This section collects comparison measures used in the process-mining literature and writes each in the $(r,\mu)$ form of Section~\ref{sec:kernel-calculus}. It is the Stage~2 filter, in which a question is checked against a map's kernel and the comparisons whose kernels cover it are ruled out. The supplementary material tabulates every row, derives it, and proves the results quoted here.

Classical process comparison is not stochastic, comparing languages and structures without a
probability measure. It is the same geometry under the projection replacing each row of a chart by
the maximum-entropy law on its support, which discards the frequencies and keeps only which
continuations are possible. Under that projection $\dSMD$ becomes an overlap of supports. The block
states of $\rP$ then give the square root of the Hamming distance between block-successor tables,
and the activity pairs of $\rPi$ the square root of Kemeny's metric~\citep{KemenySnell1962} on total
orders. Only the index set changes, and with it the normaliser.

One named comparison sets a log against a model. The log's empirical stochastic language is a
point of the same space, so the comparison is a distance in this geometry with one argument
empirical.

\begin{definition}[Fit residual]
\label{def:fit-residual}
Let $p$ be a log's empirical stochastic language and $q=\rT(m)$ a model's. Write
$a=\sum_{w\notin\operatorname{supp}q}p(w)$ for the log mass the model cannot produce and
$b=\sum_{w\notin\operatorname{supp}p}q(w)$ for the model mass never observed. The residual
is the pair $(a,b)$ together with $\dBA(p,q)$.
\end{definition}

At the level of supports the two masses are the quantities process mining names fitness and
precision, $a$ being observed behaviour the model does not admit and $b$ admitted behaviour never
observed. They are what the projection above retains. The angle carries the rest, splitting into
a floor forced by the support mismatch and a remainder that is disagreement about frequencies. The
residual depends on the model only through $\rT(m)$, so trace mass removed as noise by a pre-fit
filter lands in $a$ instead of being explained.

The angle carries a null of its own. Rescaled, $16N\sin^{2}(\dBA(p,q)/4)$ is the
power-divergence statistic of~\citet{cressieMultinomialGoodnessoffitTests1984} at
$\lambda=-\tfrac12$, a family containing Pearson's $X^{2}$ and the likelihood-ratio $G^{2}$ and
sharing one $\chi^{2}$ null distribution. A correct model does not score zero, since a finite
sample differs from the law generating it. The reference used throughout is therefore a parametric
bootstrap from the fitted model at the same $N$. That puts the null on the angle's own scale and
holds at boundary weights where the $\chi^{2}$ asymptotics do not. The supplementary material
derives the identity.

The identity chart $\rN(e_{D})=e_{D}$ returns the model vector of
Equation~\eqref{eq:model-vector} and $\tile$ is injective by Proposition~\ref{prop:canonical-tiling}, so
$\rN$ and $\rB$ have kernel $\{0\}$ and pull $\dBA$ back to metrics. At faithful memory limit $1$
(Theorem~\ref{thm:faithful}) the support pattern of the block chart $\rP$ is the footprint
construction of~\citet{aalstProcessMiningData2016} computed on blocks rather than activities,
distinct from the directly-follows graph of Theorem~\ref{thm:dfg-rank}.

Two families sit neither on the boundary nor below it. The activity footprint $\rF$ of the
$\alpha$-algorithm~\citep{aalstProcessMiningData2016} records which of four relations holds for each
activity pair, and compares two such records under $\dSMD$. Its third relation is incomparability in
the poset, so $\rF$ separates the concurrency of Example~\ref{ex:block-matrix} from the balanced
interleaving, which no chart factoring through $\rT$ does. It identifies other pairs instead,
so $\ker\rF$ and $\ker\rT$ are incomparable, $\rF$ is not injective, and the pullback is a
pseudometric. The footprint $\rF$ itself is not trace-observable, though the $\alpha$-algorithm's estimator of it
is, declaring two activities concurrent when the flat log contains both adjacencies
(Section~\ref{sec:exp-synthea}). The cophenetic process-tree distance~\citep{sanchezcharlesProcessModelComparison2016}
and the block-tree distance of~\citet{baeProcessMiningMeasuring2006} are confined to a series-parallel
decomposition. A process tree has no node for a prime, so on a poset containing a prime the
cophenetic chart is undefined, and forcing a model into a tree is a coarsening with non-trivial
kernel, hence a pseudometric. On series-parallel models with distinct labels the cophenetic distance is a metric.

Below the boundary Theorem~\ref{thm:kerL-linrel} characterises every kernel in advance, and only
metric status is left open. Five comparisons factor through $\rT$. They are the Jensen--Shannon
distance for stochastic conformance~\citep{liJensenShannonDistance2025}, earth-mover stochastic
conformance~\citep{leemansStochasticProcessMining2021}, the pairwise chart $\rPi$ with Kemeny's order
metric~\citep{KemenySnell1962,azziniNewApproachIdentifying2020} as its point-mass limit, behavioural
profiles~\citep{weidlichEfficientConsistencyMeasurement2011,kunzeBehavioralSimilarityProper2011}
compared under the Jaccard set coefficient, and the activity bag $\rL$. Each therefore inherits
$\ker\rT$ and each pullback is a pseudometric, with $\ker\rT\subsetneq\ker\rPi$
(Section~\ref{sec:monotone}, with a witness to strictness in the supplementary material), a profile factoring through $\rPi$, and $\ker\rL$ incomparable with
$\ker\rPi$. For the earth-mover row the ground cost decides metric status alone, since the kernel is
$\ker\rT$ for any ground cost separating distinct traces, and Example~\ref{ex:block-matrix} sits at distance zero under any
normalisation of the Jensen--Shannon divergence.

A block-frequency marginal of the tiling is left untabulated for want of a comparator. The
nearest alternatives, the performance spectrum~\citep{denisovUnbiasedFineGrainedDescription2018} and
queue mining, measure timing and throughput, not block frequency. Comparisons the dictionary does not
cover are discussed in Section~\ref{sec:discussion}.

\section{Estimation from data}
\label{sec:data}

Traditionally, a model discovered from a log arrives as a variant family without weights, so the weights $\rho$ are estimated from the log and $r(m)$ follows from a chart fixed before any data. Definition~\ref{def:stochastic-language-map} fixes $\rho(\cdot\mid D)$ from the family and the scheduling law, and collecting these as the columns of the \emph{law matrix} $A^{\nu}$ gives a weight vector $\rho$ the trace marginal $A^{\nu}\rho$ of Equation~\eqref{eq:marginal}. Each execution draws $D$ with chance $\rho(D)$ and then a linear extension under $\nu_{D}$, so a trace $w$ has probability $(A^{\nu}\rho)(w)$, and on an i.i.d.\ sample of $N$ traces maximum likelihood over the weights alone gives
\begin{equation}
\label{eq:weight-mle}
  \hat\rho_{N}\;=\;\operatorname*{arg\,max}_{\rho\in\simplexN}\;
  \sum_{i=1}^{N}\log\,(A^{\nu}\rho)(w_{i}).
\end{equation}
A sum of logarithms of linear functions is concave, so the problem is convex whichever law builds $A^{\nu}$, and such a mixture is identifiable exactly when its component laws are linearly independent~\citep{yakowitzIdentifiabilityFiniteMixtures1968}, which is full column rank of $A^{\nu}$.

Where the variant supports are disjoint the maximiser is instead unique and available in closed form. If every observed trace lies in one of the supports, the log-likelihood separates into a term in $\rho$ and a term free of it, making $\hat\rho_{N}(D)$ the proportion of traces falling in $\ell(\lin{D})$, obtained in one pass over the log.

Only the empirical stochastic language enters~\eqref{eq:weight-mle}, the multiset of traces and nothing else. Timestamps, resources and durations play no part, and neither does the within-block scheduling law, which cancels in the block chart.

The block chart $\rP$ is formed from $\hat\rho_{N}$ by substitution. The context tree is fixed across the comparison, selected from the compared models jointly and never from a sample's own weights, so the state space is not random and the matrix entries are ratios of linear forms in the weights, as the continuity result below needs.

\begin{proposition}[Plug-in consistency]
\label{prop:chart-consistency}
Fix any scheduling law satisfying Assumption~\ref{ass:fixed-law}, let the observed executions be independent draws from the marginal~\eqref{eq:marginal}, and suppose the variant family is identifiable, the law matrix $A^{\nu}$ having full column rank. Then, for every chart $r$ continuous in the weights and every metric $\mu$ continuous on its image,
\begin{itemize}
\item the maximum-likelihood weights of Equation~\eqref{eq:weight-mle} converge,
      $\hat\rho_{N}\xrightarrow{\text{a.s.}}\rho$, everywhere on the simplex, and the true weights need not lie in its interior.
\item the chart of them converges, $r(\hat\rho_{N})\xrightarrow{\text{a.s.}}r(\rho)$.
\item the distance between two of them converges,
      $\mu\bigl(r(\hat\rho_{N}),r(\hat\rho'_{N})\bigr)\xrightarrow{\text{a.s.}}\mu\bigl(r(\rho),r(\rho')\bigr)$.
\end{itemize}
\end{proposition}

The law enters only through the columns of $A^{\nu}$, so the conclusion holds for every law admitted by Assumption~\ref{ass:fixed-law}, and Assumption~\ref{ass:uniform} selects which matrix is meant, and with it the limit converged to. Where a block marginal vanishes $\rP$ is discontinuous. The convention of Section~\ref{sec:block-chart} pins the row of an unvisited state, so the distance remains computable and only the plug-in limit lapses.

Independence of the executions is a property of the execution notion and not of the log, and object-centric event data supplies an execution only once that notion is chosen \citep{adamsDefiningCasesVariants2022}. Independence holds where the executions so obtained are disjoint and fails in two ways otherwise. An event related to several objects of one type is duplicated once per object, size-biasing the empirical distribution over executions. An object outliving a single execution instead merges the log into one component, reducing the number of independent draws whatever the nominal execution count, and no reweighting repairs it. Events belonging to no execution require a third declaration, retained as an insertion channel that Theorem~\ref{thm:monotone} bounds, or discarded, which reweights the variants unequally. All three are settled at the representation alongside the chart (Section~\ref{sec:discussion}).

\begin{definition}[Estimability]
\label{def:estimability}
For a log grouped into executions and a variant family $\Nset$ carrying the law matrix $A^{\nu}$, the \emph{estimability} of the family is $\sigma_{\min}(A^{\nu})$, the smallest singular value of the law matrix.
\end{definition}

Estimability reports one way Proposition~\ref{prop:chart-consistency} can fail, a rank failure, and is computed from the variant family before any distance. The value $\sigma_{\min}$ is the smallest change in the trace marginal that a unit change in the weights can produce, so it answers two questions at once. At zero it is a verdict. Some direction in weight space alters no trace probability, which is the rank deficiency of the linearisation relation of Theorem~\ref{thm:kerL-linrel}, and there no chart of the weights is estimable at any sample size. A rank-deficient family can fit the log exactly, every maximiser being a global maximum, so fitness and precision are perfect where the weights are arbitrary.

Away from zero it is a scale, the conditioning of the law matrix, and it enters the rate below as $\sigma_{\min}^{-1}$, so halving it quadruples the log needed for a fixed accuracy. The columns are probability vectors, placing $\sigma_{\min}$ in $[0,1]$. A block of $j$ concurrent activities spreads its single law over $j!$ interleavings and gives $\sigma_{\min}=(j!)^{-\frac{1}{2}}$, so observing the order in place of the poset multiplies the sample size needed for a fixed accuracy by $w!$, while the linearly dependent laws of Example~\ref{ex:block-matrix} give $\sigma_{\min}=0$ for every $\theta$.

The rate is the standard one for a finite mixture with known components, holding everywhere on the simplex, the boundary included, $\lVert\hat\rho_{N}-\rho\rVert=O_{P}(\sigma_{\min}(A^{\nu})^{-1}N^{-1/2})$~\citep[Thm.~5.52]{vandervaartAsymptoticStatistics1998}, and a locally Lipschitz chart and metric carry the order. The supplementary material derives it and gives the standard error. With $\nu$ free the law matrix depends on the parameter and~\eqref{eq:weight-mle} becomes a sum of logarithms of a form bilinear in $(\rho,\nu)$, concave in each argument alone and in neither jointly, so weights and law trade against each other. The energy tilt $\nu^{\vartheta}$ of Section~\ref{sec:stochastic-language-map} is the tractable case, one parameter profiled out by a search over $\vartheta$ with a concave problem in $\rho$ at each value.

\section{Experiments}
\label{sec:experiments}

We use two simulated settings, one synthetic and one from a clinical patient simulator, each built so that a failure of recovery is attributable to the chart or to the geometry. Throughout we cluster by average linkage and score recovery by the adjusted Rand index, which reads $1$ at exact recovery and $0$ at chance. Comparisons use the geometry of Section~\ref{sec:geometry}, $\dBA$ on stochastic languages and the $1/\sqrt{\lvert X\rvert}$-normalised $\dSMD$ on block matrices.

\begin{figure}[!b]
\centering
\includegraphics[width=\columnwidth]{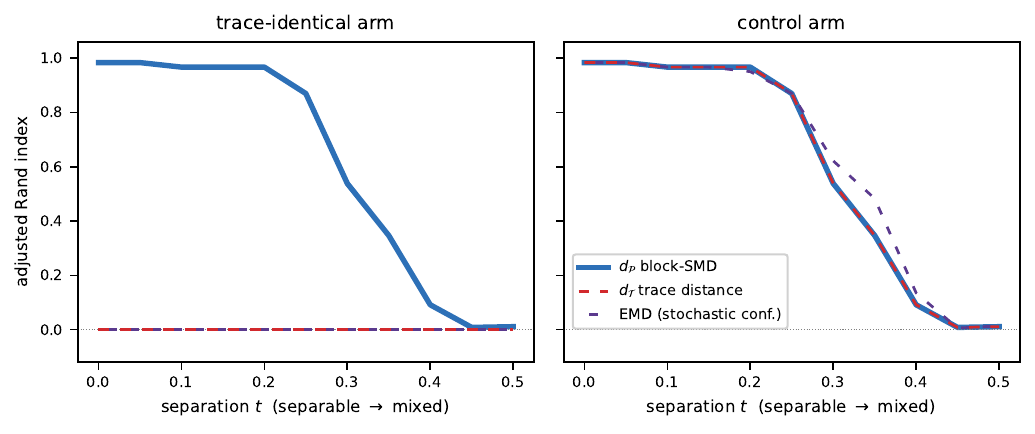}
\caption{Clustering ARI against the cluster-spacing parameter $t$, from far apart at $t=0$ to
coincident at $t=\tfrac12$, for the block-chart distance $\dP$, the trace distance $\dT$ and the
earth-mover distance, each point a mean over ten fleets of twenty-four models.
\emph{Left, the trace-identical arm.} Both trace-based curves are flat at chance across the
sweep, as Proposition~\ref{prop:log-blind} forces, while $\dP$ starts at near-perfect recovery
and degrades smoothly. \emph{Right, the control arm, a positive control.} The mixture's stochastic language varies
with the mixing weight and all three curves are qualitatively similar. On this one-parameter
family $\dT$ is a fixed multiple of $\dP$, so the dashed $\dT$ curve is drawn over the solid one.}
\label{fig:graceful}
\end{figure}

\subsection{Cluster recovery with and without a trace-level signal}
\label{sec:exp-graceful}

The first experiment asks which comparisons recover two groups of sites differing in structure, and how that recovery decays as the groups are brought together. Each site is a convex mixture with weight $p\in[0,1]$,
\begin{equation}
m(p)=(1-p)\,m_A+p\,m_B .
\label{eq:morph}
\end{equation}
The first component is the concurrent block $m_A=e_{a;(b\otimes c)}$. Two second components stand in turn for $m_B$ and define the two arms. The trace-identical arm takes the balanced interleaving $\mBi=\tfrac12\,e_{a;b;c}+\tfrac12\,e_{a;c;b}$, whose order a log records indistinguishably from one that $m_A$ generates. The control arm takes the exclusive choice $\mBe=\tfrac12\,e_{a;b}+\tfrac12\,e_{a;c}$, which drops one activity from every trace.

\begin{proposition}[Stochastic language in the mixing weight]
\label{prop:log-blind}
Take the family~\eqref{eq:morph}. With $m_B=\mBi$ the two components share a stochastic language, $\rT(\mBi)=\rT(m_A)$, and it is constant in the mixing weight,
\begin{equation}
\rT(m(p))=\rT(m_A)=\tfrac12\,e_{abc}+\tfrac12\,e_{acb}\quad\text{for every } p .
\label{eq:ident-lang}
\end{equation}
With $m_B=\mBe$ they differ, $\rT(\mBe)=\tfrac12\,e_{ab}+\tfrac12\,e_{ac}$, and linearity of $\rT$ makes the language vary with that weight,
\begin{equation}
\rT(m(p))=(1-p)\,\rT(m_A)+p\,\rT(\mBe).
\label{eq:diff-lang}
\end{equation}
Hence in the first arm any comparison factoring through $\rT$, and any process-discovery algorithm whose only input is the flat log, returns the same output on every model of the family.
\end{proposition}

Experimentally, a fleet is twenty-four sites, twelve drawn about each of two centres in the mixing weight, the centres sitting at $\tfrac12\mp0.4(1-2t)$ under a cluster-spacing parameter $t\in[0,\tfrac12]$. At $t=0$ one group is peaked near $m_A$ and the other near $m_B$, so a comparison that separates the two components recovers the groups, and at $t=\tfrac12$ both centres coincide at the balanced mixture, the groups are drawn from one distribution, and chance is the best any comparison can return. Figure~\ref{fig:graceful} plots ARI against $t$ for the block-chart distance $\dP$, the trace distance $\dT$, and the earth-mover distance between stochastic languages (EMD), the standard stochastic-conformance metric. In the trace-identical arm $\dP$ falls from ARI $0.98$ at $t=0$ to $0.01$ at $t=\tfrac12$, while both trace-based curves sit at $0$ throughout. The exhibit is computed on exact model vectors, so those two curves are $0$ by algebra. In the sampled regime, a separate measurement, every miner run on logs from this fleet stays inside a chance band of $[-0.04,0.11]$, with the six miners, the five of Section~\ref{sec:exp-synthea} together with Alpha+, and the null behind that band given in the supplementary material. In the control arm the poles differ in activity bag as well as order, so all three recover. That arm is a positive control. The trace-based comparisons separate sites whenever the traces carry the difference, so their flat curves in the first arm are attributable to the representation alone.

Cluster quality is not an indicator of real structure. We process three fleets by the standard route, a structureless fleet together with the two arms above. Each site contributes an empirical trace language, compared by trace distance, and each returned partition is scored by its silhouette, the internal signal a practitioner reads to judge a clustering. On a fleet with \emph{no} grouping that route returns a two-cluster partition at silhouette $0.580$, inside the range conventionally read as reasonable structure~\citep{rousseeuwSilhouettesGraphicalAid1987,kaufmanFindingGroupsData1990,Akta__2024}. On a fleet whose two genuine groups the flat log conflates (Proposition~\ref{prop:log-blind}), it returns silhouette $0.575$ at ARI $0.014$, and on a third fleet, where the difference does reach the log, $0.911$. All three sit in the band associated with structure, so the silhouette approach reports all three as real, with no external warning. One remedy, afforded only where a resampling pipeline is available, generates a second independent log per setting. That second log agrees with the first at chance on the first two fleets and exactly on the third, and ten times the log size does not change the result. The three fleets come from fixed-seed scripts available from the authors on request.

\subsection{A simulated multi-site clinical fleet}
\label{sec:exp-synthea}

Scoring recovery on real data needs three things at once, a known generating model, a site key by which a fleet can be assembled, and structure that a flat log does not record. The third is \emph{within-execution} concurrency, an execution whose poset is not a chain, seen in a log for example as a shared timestamp or as a pair of overlapping durations. In a corpus screen of $59$ openly available or credentialed-access candidate logs, none supplies all three, most failing on the site key alone. In BPI Challenge 2012~\citep{vandongenBPIChallenge20122012}, $99.6\%$ of the within-execution tie blocks are one resource emitting several activities at one recorded millisecond, and once the system actor is excluded only two of the $11\,361$ blocks involve more than one, so the ties record batched logging of serial work. BPI Challenge 2013~\citep{steemanBPIChallenge20132013} supplies a usable site key and records no within-execution tie at any granularity. The clinical interval data that does carry the signal, the medication administrations of the credentialed-access eICU Collaborative Research Database~\citep{pollardEICUCollaborativeResearch2018,pollardEICUCollaborativeResearch2019,pollardPhysioNetGlobalPlatform2026}, supplies a real hospital fleet key and no known model against which recovery could be scored. We therefore construct a controlled experimental fleet.

These figures do not compose into a single prevalence. Concurrency in a log is declared by a rule and not read off the data, so the share of executions carrying it moves with the rule. The same corpus returns no within-execution structure at any granularity under a shared-timestamp rule on BPI Challenge 2013, batched serial work under that rule on BPI Challenge 2012, and overlapping durations by construction under an interval rule on eICU. The disagreement is a property of the screen's subject rather than a defect of the screen. A log held as a multiset of traces records no concurrency to find, so any figure reported for its prevalence belongs to the declaring rule together with the instrumentation that fed it. That rule is the first of the representational decisions this paper makes explicit, and it is taken before any distance is computed.

\begin{figure*}[t]
\centering
\ifdim\columnwidth<\textwidth
\includegraphics[width=0.825\textwidth]{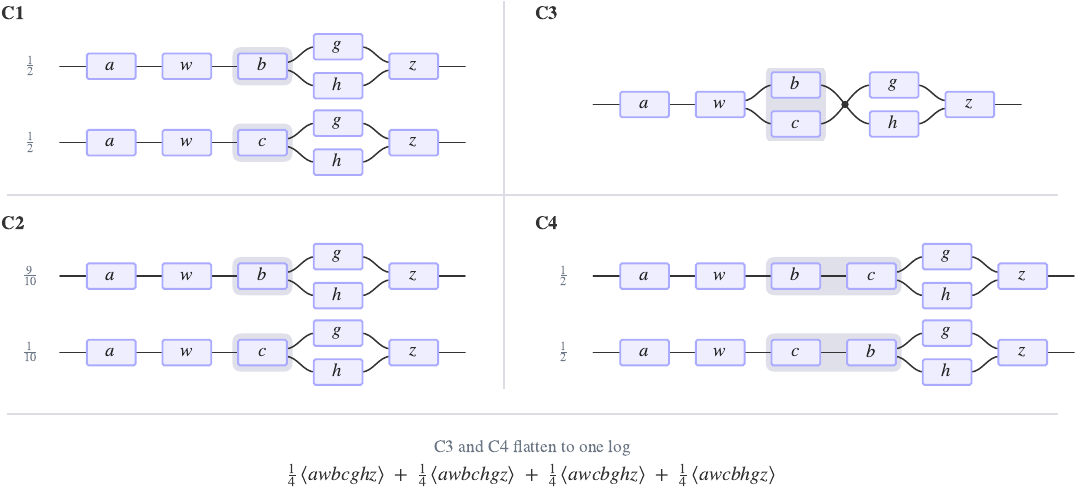}%
\else
\includegraphics[width=\textwidth]{figures/fig-mechanism.pdf}%
\fi
\caption{The four generating modules, one whole model to a quadrant, and the log C3 and C4
share. The tint marks the stretch over which the four differ, and a module emitting more than
one variant has them stacked at their weights. C1 and C2 emit one of the diagnostic pair and
differ only in the odds, C3 emits both at one timestamp and unordered, and C4 emits both at
two, so a per-execution order is recorded. The dot in C3 is a junction of wires, every activity
entering it preceding every activity leaving it. Flattening shuffles each co-timestamped block
uniformly, so C3 and C4 flatten to one trace distribution and a discovery algorithm reading
only the flattened log returns one model for both.}
\label{fig:mechanism}
\end{figure*}

Twenty sites come from the synthetic patient simulator Synthea~\citep{walonoskiSyntheaApproachMethod2018} on four purpose-built generative modules, one per ground-truth cluster, five sites per cluster differing only in their simulated patients (generation details in the supplementary material). Every module emits one ambulatory encounter over the same skeleton, an arrival $a$, a triage $w$, the diagnostic pair $b,c$, a further concurrent block $g,h$ and a discharge $z$, differing only in how $b$ and $c$ are wired into it (Figure~\ref{fig:mechanism}). The modules form two pairs, each varying one setting and holding the other fixed, so a difference in outcome has a single cause. C1 and C2 emit exactly one of the pair and differ only in the odds on it, $0.5/0.5$ against $0.9/0.1$, a difference in branch probability alone. C3 and C4 emit both, concurrently in C3 ($b\otimes c$, one shared timestamp) and in a per-execution choice of orders at balanced weight in C4 ($b;c$ or $c;b$, two distinct timestamps), a difference in concurrency alone, the trace-identical arm of Section~\ref{sec:exp-graceful} instantiated in a clinical fleet. The block $g\otimes h$ is concurrent in every module, so the presence of concurrency separates nothing. Activity codes are synthetic and the topologies stylised (12-lead ECG $b$, troponin draw $c$).

Each execution becomes a model by two routes differing only in their treatment of co-timestamped events. Route~A, the standard pipeline, flattens the execution to a total order by shuffling co-timestamped events uniformly, then discovers a model with each of five miners, Alpha, Heuristics, Inductive, integer linear programming (ILP) and Split Miner, whose variant weights $\rho(D)$ are estimated by~\eqref{eq:weight-mle}. Every discovered family here has disjoint variant supports, so the closed form of Section~\ref{sec:data} attains the maximum exactly and the fit residual of Definition~\ref{def:fit-residual} is consistent with its parametric-bootstrap null, two sites of twenty lying above their $95$th percentile where one is expected. The limit on Route~A is a property of what its representation keeps, independent of how well its weights were fitted. Route~B keeps the timestamps, reading the partial order the data already carries,
\begin{equation}
x<y \iff \operatorname{time}(x)<\operatorname{time}(y),
\label{eq:tsorder}
\end{equation}
with co-timestamped events incomparable, a parallel block. A clock quantises, so co-timestamped means within one tick of the log's own recorded resolution, and that resolution is the first thing the corpus screen measures. The poset follows the fixed rule~\eqref{eq:tsorder} without a search, so $\rho(D)$ is exact by counting. Proposition~\ref{prop:log-blind} constrains comparisons factoring through $\rT$, whose values are distributions over traces, and~\eqref{eq:tsorder} reads a timestamp, a datum the trace does not carry. The trace-identical arm's two components have one trace distribution, so a statistic separating them is not a function of the traces alone, and Route~A discards the timestamps as its first step. Recovery under Route~B therefore locates the loss in the representation rather than in the data, and the rule declaring which events are concurrent carries a kernel of its own (Section~\ref{sec:discussion}). Both routes land in the same chart, the block-transition matrix $\rP^{(1)}=\Phik[1]\circ\rB$, differing only in the concurrency relation. The discovered model is non-stochastic, both routes' weights are empirical frequencies, and the comparison never uses a miner's own routing probabilities.

Reading the observed partial order recovers all four clusters at ARI $1.000$, while the best total-order route, the flat log itself, merges C3 with C4 and stops at $0.633$. The cluster count is supplied, not selected, so at four clusters a route merging C3 with C4 must split a cluster it does recover. Write $d(\text{C3},\text{C4})$ for the distance between the C3 and C4 site models in a route's own chart and geometry, set against that route's within-cluster spread, the noise scale a between-cluster gap has to stand clear of. Under Route~A the Inductive Miner infers $b\otimes c$ for C4 as for C3, correctly, their flattened logs being identical, so the discovered signature offers only that variant, the weights are forced onto it, and the two clusters receive one matrix. The distance $d(\text{C3},\text{C4})$ is zero identically, at every seed. The two total-order routes with a nonzero value, the flat log and Split Miner, exceed their own within-C3 spread by an amount comparable to the across-realisation variation of those quantities, so neither margin reliably differentiates the clusters. All five routes and both discriminating pairs are tabulated in the fleet-recovery table of the supplementary material.

\begin{figure*}[t]
\centering
\includegraphics[width=0.72\textwidth]{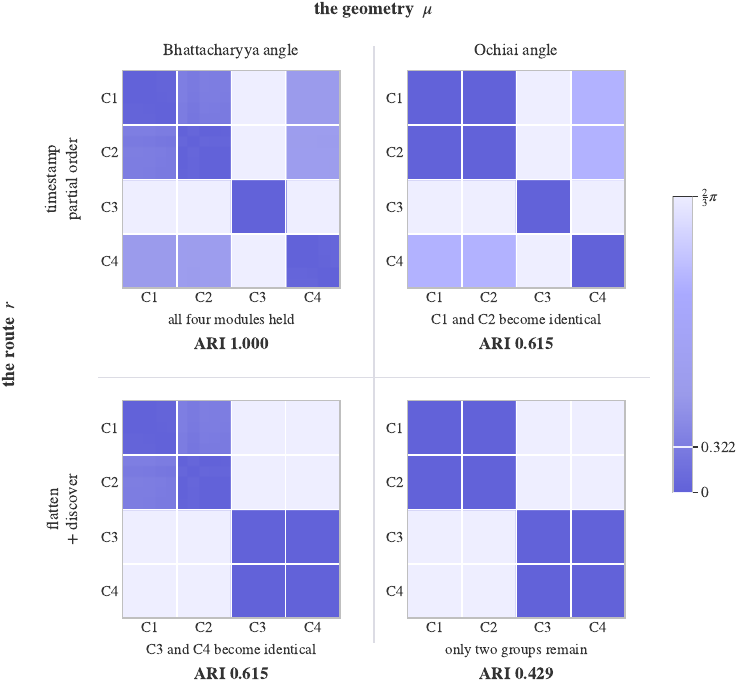}
\caption{One fleet, two routes, two geometries. The twenty sites' pairwise distances under
each pairing of route $r$ and geometry $\mu$, the sites ordered by generating module and every
panel on one scale, both geometries being angles bounded by $\pi$. Dark is near. Flattening
drives C3--C4 to zero, the support-only Ochiai angle drives C1--C2 to zero, and the observed
partial order under the Bhattacharyya angle holds all four modules apart.}
\label{fig:ablation}
\end{figure*}

Degrading either the representation or the geometry alone merges two of the four clusters, and on this fleet the two losses are the same size (Figure~\ref{fig:ablation}). With the representation held at the observed partial order, reducing $\mu$ to the support-only comparison drops recovery from ARI $1.000$ to $0.615$, C1 and C2 merging under that swap as C3 and C4 do under a total order. Flattening the representation as well leaves a distance matrix with two distinct values, and so two groups, at ARI $0.429$. On the C3--C4 pair, whose rows are point masses, the support-only comparison returns what the full comparison returns, both $\pi\sqrt{4/9}$, the square-root Hamming reading of Section~\ref{sec:dictionary}, so that pair tests the representation alone. On the C1--C2 pair it returns exactly $0$ where the block chart returns $0.322\pm0.026$ (mean $\pm$ s.d. over the twenty-five C1--C2 site pairs), so there the representation is inert and the geometry decisive. The fleet establishes that the geometry must compare probabilities and not support, though not that the Bhattacharyya angle is the best such geometry, any $\mu$ separating the branch row recovering C1 from C2 here. The distances and their closed forms are in the supplementary material.

After flattening, the only difference left between C3 and C4 is how often $b$ is recorded before $c$. Flattening shuffles each tied pair uniformly, so it writes $b$ first in half of C3's executions, while in C4 that same frequency is the module's own order weight. The two logs coincide where shuffle and weight agree, here at C4's balanced $\tfrac12$. Moving C4's weight off $\tfrac12$ restores the difference once the move exceeds sampling noise. That leaves a band of coincidence on the weight axis, narrowing as $N^{-1/2}$ and measured in the supplementary material. A tilted shuffle instead relocates the coincidence to whichever weight matches the tilt rather than removing it. Route~B reads no such frequency and returns $\dSMD=2.094$ at every weight, tilt and execution count of the sweep, reported in the supplementary material.

One total-order route, the flat log itself, does return a nonzero $d(\text{C3},\text{C4})=0.183$ where the other discovered-model routes return zero, Split Miner's nonzero value being site-to-site variation in the model it discovers. That margin is not evidence that it keeps more structure. Theorem~\ref{thm:monotone} compares two charts of one model, whereas here the two sides are different models, the log's own variant distribution and a model rebuilt from that log by discovery and weight fitting. No stochastic map carries one to the other, so the $0.183$ measures the rebuilding. The loss under a total order is specific to concurrency. Flattening destroys the order inside a tie and leaves the branch frequencies intact, so Route~A still recovers C1's and C2's weights from the discovered model ($\hat\rho_N\approx0.5/0.5$ against $0.9/0.1$) and holds that pair apart. Over the twenty-five C1--C2 site pairs the mean $\dSMD$ is $0.322\pm0.026$, against a mean of $0.020$ over the ten pairs within C1, and the closest C1--C2 pair, at $0.260$, stands clear of the widest pair within either cluster, at $0.060$.

Clustering via a model is also sensitive to ties. Perturbing each co-timestamp independently inside the execution's own smallest inter-event gap, so that the macroscopic order survives and only the ties are broken, serialises every tied block in the fleet. The distance $d(\text{C3},\text{C4})$ then falls from $2.094$ to $0.077\pm0.030$ against a within-C3 spread of $0.006\pm0.002$, a between-cluster gap surviving in sign but not in magnitude. The ARI of $1.000$ is therefore a claim about tie-preserving clocks and not about arbitrary timestamp noise.

Two checks close the fleet. The chart used above reads one block of history when deciding what follows, while Theorem~\ref{thm:faithful} certifies this fleet only from two blocks onward, $\kstar=2$, so that chart is a truncation of the faithful one. Repeating the clustering at $k=2$ and $k=3$ returns the same grouping, an agreement verified and not assumed. The second check runs on the variant family that separating C3 from C4 requires, before any distance is computed. Estimability (Definition~\ref{def:estimability}) vanishes on exactly the two concurrency-versus-choice families, this fleet's C3--C4 pair and the trace-identical arm of Section~\ref{sec:exp-graceful}, and is bounded away from zero on the two controls. In both cases the vanishing is a property of the representation, the sample being irrelevant to it, so no larger log fixes it. The families the miners discovered are instead full rank, so the log determined their weights but not their concurrency.

\section{Discussion and conclusions}
\label{sec:discussion}

A distance between care pathways reports those differences its representation preserves, and no others. Equation~\eqref{eq:pullback} states the dependence, $\mu$ grading differences and $\ker r$ deciding which differences exist. That kernel is concrete in five ways. It has an explicit basis and a membership test at the trace level (Theorem~\ref{thm:kerL-linrel}). The graph most tools compute first has a rank against it (Theorem~\ref{thm:dfg-rank}). Charts are ordered by how much each discards (Theorem~\ref{thm:monotone}). A construction ends the search for a faithful chart (Theorem~\ref{thm:faithful}). A certificate is computed from the variant family before any distance is taken (Definition~\ref{def:estimability}). The choice of comparison thereby moves from the tool to the decision.

The experiments give the dependence its magnitude. Flattening the observed partial order inside the block chart, the geometry held fixed, drops cluster recovery on the clinical fleet from an ARI of $1.000$ to $0.615$. Holding the representation and replacing the Bhattacharyya angle by a support-only comparison drops it to the same $0.615$. Each loss is a kernel, so no larger sample recovers either. The same accounting applies to the instrument. Concurrency in a log is declared and not observed. Two events count as concurrent because both orders occur, because their timestamps agree (Equation~\eqref{eq:tsorder}), or because their intervals overlap, and each rule is a channel with a kernel of its own.

The frame locates three neighbouring literatures. The practitioner catalogue of directly-follows misbehaviour \citep{vanderaalstPractitionersGuideProcess2019} contracts to a rank. For a fixed multiset of activities, every table the graph reports is induced also by a model with no concurrency, so a reported concurrency originates in the miner's rule (Section~\ref{sec:dfg}). The stochastically known log programme \citep{galEverythingThereKnow2023} coincides with this frame on $\ker\rT$. At the timestamp, jitter breaking a genuine tie acts upstream of $r$ as one more linearisation relation. Comparisons returning an interpretable difference or a test verdict \citep{armas-cervantesBehavioralComparisonProcess2014,garcia-banuelosCompleteInterpretableConformance2018,nguyenMultiperspectiveComparisonBusiness2018} pull back no $\mu$, so the dictionary does not tabulate them.

The evaluation is simulated throughout. The corpus screen of Section~\ref{sec:exp-synthea} found no accessible public log posing the comparison under a known generating model, so the ground truth was constructed and the figures describe a fleet whose structure we fixed. They also assume tie-preserving instrumentation. Perturbing each co-timestamp inside its execution's smallest inter-event gap serialises every tied block, and the C3--C4 gap survives in sign but not in magnitude. Two conditions bound the framework itself. Directed scores, partial-order alignment costs among them \citep{luConformanceCheckingBased2015}, keep $\ker r$ and the coarsening bound \citep{amariInformationGeometry2021} while losing the triangle inequality on which the normaliser of Section~\ref{sec:geometry} rests, and Proposition~\ref{prop:chart-consistency} requires the executions to be independent draws, which object sharing denies. The comparison pipeline, with the candidate list behind the screen, is implemented in the open-source \texttt{proc-posets} package (\url{https://github.com/AntonyRLee/proc-posets}), archived at \url{https://doi.org/10.5281/zenodo.22757993}.

\subsection{Implications for practice}

The three decisions of Section~\ref{sec:intro} divide along the trace-level boundary. For the lead grouping several hospitals, Stage~2 excludes every trace-level chart before a log is acquired. No fitness, entropy-based precision \citep{polyvyanyyMonotonePrecisionRecall2020}, entropic relevance \citep{alkhammashEntropicRelevanceMechanism2022} or earth-mover score separates models differing only in what the boundary conflates. For the manager auditing one site, a change confined to whether two steps run concurrently is reported by the block chart and absent from the flattened log (Section~\ref{sec:exp-synthea}). For the commissioner benchmarking two providers, the faithful representation is also the less expensive to compute, and the choice falls to Stage~2. The execution notion is a Stage~2 decision as well. Every recorded attribute induces a candidate grouping on the charts of Corollary~\ref{cor:object-centric}, each scored by Definition~\ref{def:estimability} before any distance, so competing definitions \citep{adamsDefiningCasesVariants2022} are weighed on an equal footing.

Whether some invariant of the order alone decides equality of stochastic languages is open, as is whether any log-derived channel is injective outright. Both are relative to the scheduling law, since under a tilted law the proportions of the interleavings carry signal. The intermediate dependence regime, each execution sharing objects with boundedly many others, has no asymptotic statement here. A public log recording genuine concurrency under a known generating model would turn the constructed fleet into a test.

Which differences a distance can register is settled by the representation the comparison factors through, before any log is opened. The paper makes that settlement checkable, the kernel computed, the charts ordered, the log certified. A similarity score reported with its kernel is evidence for a decision. Reported without it, the score is a number.

Every result a trace-level comparison has reported remains valid within its resolution. The resolution states which questions were in range, and identifies, before a study is commissioned, when a richer representation is needed and when the cheaper one suffices. Where the question concerns what runs at the same time, the answer is not present in the sequence, and no analysis of the sequence recovers it.

\section*{Declarations}

\noindent\textbf{CRediT authorship contribution statement.}
\textbf{Antony R. Lee:} Conceptualisation, Methodology, Software, Formal analysis, Investigation,
Validation, Visualisation, Writing -- original draft, Writing -- review \& editing.
\textbf{Peter Ti\v{n}o:} Supervision, Writing -- review \& editing. \textbf{Iain B. Styles:}
Supervision, Writing -- review \& editing.

\noindent\textbf{Competing interests.} The authors declare that they have no known competing
financial interests or personal relationships that could have appeared to influence the work
reported in this paper.

\noindent\textbf{Funding.} ARL acknowledges the receipt of studentship awards from the Health Data
Research UK-The Alan Turing Institute Wellcome PhD Programme in Health Data Science (Grant Ref:
218529/Z/19/Z). P.~Ti\v{n}o was supported by the EPSRC Prosperity Partnerships grant ARCANE,
EP/X025454/1.

\noindent\textbf{Data availability.} No proprietary or patient-identifiable data were used. The eICU Collaborative Research Database, examined in the corpus screen, is available to credentialed users under the PhysioNet Credentialed Health Data License. The
clinical fleet of Section~\ref{sec:exp-synthea} is generated by the open-source Synthea patient
simulator from the parameters recorded in the supplementary material. The comparison
pipeline and the candidate list behind the corpus screen are distributed in the open-source
\texttt{proc-posets} package (\mbox{\url{https://github.com/AntonyRLee/proc-posets}}), archived at
\url{https://doi.org/10.5281/zenodo.22757993}. The generation modules, their seeds and the analysis scripts are available from the authors
on request.


\begin{thebibliography}{82}
\expandafter\ifx\csname natexlab\endcsname\relax\def\natexlab#1{#1}\fi
\providecommand{\url}[1]{\texttt{#1}}
\providecommand{\href}[2]{#2}
\providecommand{\path}[1]{#1}
\providecommand{\DOIprefix}{doi:}
\providecommand{\ArXivprefix}{arXiv:}
\providecommand{\URLprefix}{URL: }
\providecommand{\Pubmedprefix}{pmid:}
\providecommand{\doi}[1]{\href{http://dx.doi.org/#1}{\path{#1}}}
\providecommand{\Pubmed}[1]{\href{pmid:#1}{\path{#1}}}
\providecommand{\bibinfo}[2]{#2}
\ifx\xfnm\relax \def\xfnm[#1]{\unskip,\space#1}\fi
%Type = Inproceedings
\bibitem[{Adams et~al.(2022)Adams, Schuster, Schmitz, Schuh and {van der
  Aalst}}]{adamsDefiningCasesVariants2022}
\bibinfo{author}{Adams, J.N.}, \bibinfo{author}{Schuster, D.},
  \bibinfo{author}{Schmitz, S.}, \bibinfo{author}{Schuh, G.},
  \bibinfo{author}{{van der Aalst}, W.M.P.}, \bibinfo{year}{2022}.
\newblock \bibinfo{title}{Defining {{Cases}} and {{Variants}} for
  {{Object-Centric Event Data}}}, in: \bibinfo{booktitle}{2022 4th
  {{International Conference}} on {{Process Mining}} ({{ICPM}})},
  \bibinfo{publisher}{IEEE}. pp. \bibinfo{pages}{128--135}.
\newblock \DOIprefix\doi{10.1109/ICPM57379.2022.9980730}.
%Type = Article
\bibitem[{Akta{\c s} et~al.(2024)Akta{\c s}, Lokman, {\.I}nkaya and
  Dejaegere}]{Akta__2024}
\bibinfo{author}{Akta{\c s}, D.}, \bibinfo{author}{Lokman, B.},
  \bibinfo{author}{{\.I}nkaya, T.}, \bibinfo{author}{Dejaegere, G.},
  \bibinfo{year}{2024}.
\newblock \bibinfo{title}{Cluster ensemble selection and consensus clustering:
  {{A}} multi-objective optimization approach}.
\newblock \bibinfo{journal}{European Journal of Operational Research}
  \bibinfo{volume}{314}, \bibinfo{pages}{1065--1077}.
\newblock \DOIprefix\doi{10.1016/j.ejor.2023.10.029}.
%Type = Article
\bibitem[{Alkhammash et~al.(2022)Alkhammash, Polyvyanyy, Moffat and
  {Garc{\'i}a-Ba{\~n}uelos}}]{alkhammashEntropicRelevanceMechanism2022}
\bibinfo{author}{Alkhammash, H.}, \bibinfo{author}{Polyvyanyy, A.},
  \bibinfo{author}{Moffat, A.}, \bibinfo{author}{{Garc{\'i}a-Ba{\~n}uelos},
  L.}, \bibinfo{year}{2022}.
\newblock \bibinfo{title}{Entropic relevance: {{A}} mechanism for measuring
  stochastic process models discovered from event data}.
\newblock \bibinfo{journal}{Information Systems} \bibinfo{volume}{107},
  \bibinfo{pages}{101922}.
\newblock \DOIprefix\doi{10.1016/j.is.2021.101922}.
%Type = Book
\bibitem[{Amari(2016)}]{amariInformationGeometryIts2016}
\bibinfo{author}{Amari, S.i.}, \bibinfo{year}{2016}.
\newblock \bibinfo{title}{Information {{Geometry}} and {{Its Applications}}}.
  volume \bibinfo{volume}{194} of \textit{\bibinfo{series}{Applied Mathematical
  Sciences}}.
\newblock \bibinfo{publisher}{Springer Japan}, \bibinfo{address}{Tokyo}.
\newblock \DOIprefix\doi{10.1007/978-4-431-55978-8}.
%Type = Article
\bibitem[{Amari(2021)}]{amariInformationGeometry2021}
\bibinfo{author}{Amari, S.i.}, \bibinfo{year}{2021}.
\newblock \bibinfo{title}{Information geometry}.
\newblock \bibinfo{journal}{Japanese Journal of Mathematics}
  \bibinfo{volume}{16}, \bibinfo{pages}{1--48}.
\newblock \DOIprefix\doi{10.1007/s11537-020-1920-5}.
%Type = Inproceedings
\bibitem[{{Armas-Cervantes} et~al.(2014){Armas-Cervantes}, Baldan, Dumas and
  {Garc{\'i}a-Ba{\~n}uelos}}]{armas-cervantesBehavioralComparisonProcess2014}
\bibinfo{author}{{Armas-Cervantes}, A.}, \bibinfo{author}{Baldan, P.},
  \bibinfo{author}{Dumas, M.}, \bibinfo{author}{{Garc{\'i}a-Ba{\~n}uelos}, L.},
  \bibinfo{year}{2014}.
\newblock \bibinfo{title}{Behavioral {{Comparison}} of {{Process Models Based}}
  on {{Canonically Reduced Event Structures}}}, in:
  \bibinfo{booktitle}{Business {{Process Management}} ({{BPM}} 2014)},
  \bibinfo{publisher}{Springer}. pp. \bibinfo{pages}{267--282}.
\newblock \DOIprefix\doi{10.1007/978-3-319-10172-9_17}.
%Type = Article
\bibitem[{Augusto et~al.(2022)Augusto, {Armas-Cervantes}, Conforti, Dumas and
  La~Rosa}]{augustoMeasuringFitnessPrecision2022}
\bibinfo{author}{Augusto, A.}, \bibinfo{author}{{Armas-Cervantes}, A.},
  \bibinfo{author}{Conforti, R.}, \bibinfo{author}{Dumas, M.},
  \bibinfo{author}{La~Rosa, M.}, \bibinfo{year}{2022}.
\newblock \bibinfo{title}{Measuring {{Fitness}} and {{Precision}} of
  {{Automatically Discovered Process Models}}: {{A Principled}} and {{Scalable
  Approach}}}.
\newblock \bibinfo{journal}{IEEE Transactions on Knowledge and Data
  Engineering} \bibinfo{volume}{34}, \bibinfo{pages}{1870--1888}.
\newblock \DOIprefix\doi{10.1109/TKDE.2020.3003258}.
%Type = Inproceedings
\bibitem[{Augusto et~al.(2018)Augusto, {Armas-Cervantes}, Conforti, Dumas,
  La~Rosa and Rei{\ss}ner}]{augustoAbstractandCompareFamilyScalable2018}
\bibinfo{author}{Augusto, A.}, \bibinfo{author}{{Armas-Cervantes}, A.},
  \bibinfo{author}{Conforti, R.}, \bibinfo{author}{Dumas, M.},
  \bibinfo{author}{La~Rosa, M.}, \bibinfo{author}{Rei{\ss}ner, D.},
  \bibinfo{year}{2018}.
\newblock \bibinfo{title}{Abstract-and-{{Compare}}: {{A Family}} of {{Scalable
  Precision Measures}} for {{Automated Process Discovery}}}, in:
  \bibinfo{booktitle}{Business {{Process Management}} ({{BPM}} 2018)},
  \bibinfo{publisher}{Springer}. pp. \bibinfo{pages}{158--175}.
\newblock \DOIprefix\doi{10.1007/978-3-319-98648-7_10}.
%Type = Article
\bibitem[{Augusto et~al.(2019)Augusto, Conforti, Dumas, {La Rosa}, Maggi,
  Marrella, Mecella and Soo}]{augustoAutomatedDiscoveryProcess2019}
\bibinfo{author}{Augusto, A.}, \bibinfo{author}{Conforti, R.},
  \bibinfo{author}{Dumas, M.}, \bibinfo{author}{{La Rosa}, M.},
  \bibinfo{author}{Maggi, F.M.}, \bibinfo{author}{Marrella, A.},
  \bibinfo{author}{Mecella, M.}, \bibinfo{author}{Soo, A.},
  \bibinfo{year}{2019}.
\newblock \bibinfo{title}{Automated {{Discovery}} of {{Process Models}} from
  {{Event Logs}}: {{Review}} and {{Benchmark}}}.
\newblock \bibinfo{journal}{IEEE Transactions on Knowledge and Data
  Engineering} \bibinfo{volume}{31}, \bibinfo{pages}{686--705}.
\newblock \DOIprefix\doi{10.1109/TKDE.2018.2841877}.
%Type = Article
\bibitem[{Azzini and Munda(2020)}]{azziniNewApproachIdentifying2020}
\bibinfo{author}{Azzini, I.}, \bibinfo{author}{Munda, G.},
  \bibinfo{year}{2020}.
\newblock \bibinfo{title}{A new approach for identifying the {{Kemeny}} median
  ranking}.
\newblock \bibinfo{journal}{European Journal of Operational Research}
  \bibinfo{volume}{281}, \bibinfo{pages}{388--401}.
\newblock \DOIprefix\doi{10.1016/j.ejor.2019.08.033}.
%Type = Inproceedings
\bibitem[{Bae et~al.(2006)Bae, Caverlee, Liu and
  Yan}]{baeProcessMiningMeasuring2006}
\bibinfo{author}{Bae, J.}, \bibinfo{author}{Caverlee, J.},
  \bibinfo{author}{Liu, L.}, \bibinfo{author}{Yan, H.}, \bibinfo{year}{2006}.
\newblock \bibinfo{title}{Process {{Mining}} by {{Measuring Process Block
  Similarity}}}, in: \bibinfo{booktitle}{Business {{Process Management
  Workshops}} ({{BPM}} 2006)}, \bibinfo{publisher}{Springer}. pp.
  \bibinfo{pages}{141--152}.
\newblock \DOIprefix\doi{10.1007/11837862_15}.
%Type = Article
\bibitem[{Beli{\"e}n et~al.(2025)Beli{\"e}n, Brailsford, Demeulemeester,
  Demirtas, Hans and Harper}]{belienFiftyYearsOperational2025}
\bibinfo{author}{Beli{\"e}n, J.}, \bibinfo{author}{Brailsford, S.},
  \bibinfo{author}{Demeulemeester, E.}, \bibinfo{author}{Demirtas, D.},
  \bibinfo{author}{Hans, E.W.}, \bibinfo{author}{Harper, P.},
  \bibinfo{year}{2025}.
\newblock \bibinfo{title}{Fifty years of operational research applied to
  healthcare}.
\newblock \bibinfo{journal}{European Journal of Operational Research}
  \bibinfo{volume}{326}, \bibinfo{pages}{189--206}.
\newblock \DOIprefix\doi{10.1016/j.ejor.2024.12.040}.
%Type = Misc
\bibitem[{Berti et~al.(2024)Berti, Koren, Adams, Park, Knopp, Graves, Rafiei,
  Liss, Unterberg, Zhang, Schwanen, Pegoraro and {van der
  Aalst}}]{bertiOCELObjectCentricEvent2024}
\bibinfo{author}{Berti, A.}, \bibinfo{author}{Koren, I.},
  \bibinfo{author}{Adams, J.N.}, \bibinfo{author}{Park, G.},
  \bibinfo{author}{Knopp, B.}, \bibinfo{author}{Graves, N.},
  \bibinfo{author}{Rafiei, M.}, \bibinfo{author}{Liss, L.},
  \bibinfo{author}{Unterberg, L.T.G.}, \bibinfo{author}{Zhang, Y.},
  \bibinfo{author}{Schwanen, C.}, \bibinfo{author}{Pegoraro, M.},
  \bibinfo{author}{{van der Aalst}, W.M.P.}, \bibinfo{year}{2024}.
\newblock \bibinfo{title}{{{OCEL}} ({{Object-Centric Event Log}}) 2.0
  {{Specification}}}.
\newblock \bibinfo{howpublished}{Preprint, arXiv:2403.01975}.
%Type = Misc
\bibitem[{Berti et~al.(2023)Berti, Montali and {van der
  Aalst}}]{bertiAdvancementsChallengesObjectCentric2023}
\bibinfo{author}{Berti, A.}, \bibinfo{author}{Montali, M.},
  \bibinfo{author}{{van der Aalst}, W.M.P.}, \bibinfo{year}{2023}.
\newblock \bibinfo{title}{Advancements and {{Challenges}} in {{Object-Centric
  Process Mining}}: {{A Systematic Literature Review}}}.
\newblock \bibinfo{howpublished}{Preprint, arXiv:2311.08795}.
\newblock \DOIprefix\doi{10.48550/arXiv.2311.08795}.
%Type = Article
\bibitem[{Boyle et~al.(2022)Boyle, Marshall and
  Mackay}]{boyleFrameworkDevelopingGeneralisable2022}
\bibinfo{author}{Boyle, L.M.}, \bibinfo{author}{Marshall, A.H.},
  \bibinfo{author}{Mackay, M.}, \bibinfo{year}{2022}.
\newblock \bibinfo{title}{A framework for developing generalisable discrete
  event simulation models of hospital emergency departments}.
\newblock \bibinfo{journal}{European Journal of Operational Research}
  \bibinfo{volume}{302}, \bibinfo{pages}{337--347}.
\newblock \DOIprefix\doi{10.1016/j.ejor.2021.12.033}.
%Type = Article
\bibitem[{Brightwell and
  Winkler(1991)}]{brightwellCountingLinearExtensions1991a}
\bibinfo{author}{Brightwell, G.}, \bibinfo{author}{Winkler, P.},
  \bibinfo{year}{1991}.
\newblock \bibinfo{title}{Counting {{Linear Extensions}}}.
\newblock \bibinfo{journal}{Order} \bibinfo{volume}{8},
  \bibinfo{pages}{225--242}.
\newblock \DOIprefix\doi{10.1007/BF00383444}.
%Type = Inproceedings
\bibitem[{Brons et~al.(2021)Brons, Scheepens and
  Fahland}]{bronsStrikingNewBalance2021}
\bibinfo{author}{Brons, D.}, \bibinfo{author}{Scheepens, R.},
  \bibinfo{author}{Fahland, D.}, \bibinfo{year}{2021}.
\newblock \bibinfo{title}{Striking a new {{Balance}} in {{Accuracy}} and
  {{Simplicity}} with the {{Probabilistic Inductive Miner}}}, in:
  \bibinfo{booktitle}{2021 3rd International Conference on Process Mining
  (ICPM)}, \bibinfo{publisher}{IEEE}. pp. \bibinfo{pages}{32--39}.
\newblock \DOIprefix\doi{10.1109/ICPM53251.2021.9576864}.
%Type = Article
\bibitem[{B{\"u}hlmann and Wyner(1999)}]{buhlmannVariableLengthMarkov1999}
\bibinfo{author}{B{\"u}hlmann, P.}, \bibinfo{author}{Wyner, A.J.},
  \bibinfo{year}{1999}.
\newblock \bibinfo{title}{Variable length {{Markov}} chains}.
\newblock \bibinfo{journal}{The Annals of Statistics} \bibinfo{volume}{27},
  \bibinfo{pages}{480--513}.
\newblock \DOIprefix\doi{10.1214/aos/1018031204}.
%Type = Book
\bibitem[{Burago et~al.(2001)Burago, Burago and
  Ivanov}]{buragoCourseMetricGeometry2001}
\bibinfo{author}{Burago, D.}, \bibinfo{author}{Burago, Y.},
  \bibinfo{author}{Ivanov, S.}, \bibinfo{year}{2001}.
\newblock \bibinfo{title}{A {{Course}} in {{Metric Geometry}}}.
\newblock Number~\bibinfo{number}{33} in \bibinfo{series}{Graduate {{Studies}}
  in {{Mathematics}}}, \bibinfo{publisher}{American Mathematical Society},
  \bibinfo{address}{Providence, RI}.
\newblock \DOIprefix\doi{10.1090/gsm/033}.
%Type = Book
\bibitem[{{\v C}encov(1982)}]{cencovStatisticalDecisionRules1982}
\bibinfo{author}{{\v C}encov, N.N.}, \bibinfo{year}{1982}.
\newblock \bibinfo{title}{Statistical Decision Rules and Optimal Inference}.
  volume~\bibinfo{volume}{53} of \textit{\bibinfo{series}{Translations of
  Mathematical Monographs}}.
\newblock \bibinfo{publisher}{American Mathematical Society},
  \bibinfo{address}{Providence, RI}.
%Type = Article
\bibitem[{Corne et~al.(2012)Corne, Dhaenens and Jourdan}]{Corne_2012}
\bibinfo{author}{Corne, D.}, \bibinfo{author}{Dhaenens, C.},
  \bibinfo{author}{Jourdan, L.}, \bibinfo{year}{2012}.
\newblock \bibinfo{title}{Synergies between operations research and data
  mining: {{The}} emerging use of multi-objective approaches}.
\newblock \bibinfo{journal}{European Journal of Operational Research}
  \bibinfo{volume}{221}, \bibinfo{pages}{469--479}.
\newblock \DOIprefix\doi{10.1016/j.ejor.2012.03.039}.
%Type = Book
\bibitem[{Cover and Thomas(2006)}]{coverElementsInformationTheory2006}
\bibinfo{author}{Cover, T.M.}, \bibinfo{author}{Thomas, J.A.},
  \bibinfo{year}{2006}.
\newblock \bibinfo{title}{Elements of {{Information Theory}}}.
\newblock \bibinfo{edition}{2nd} ed., \bibinfo{publisher}{Wiley-Interscience},
  \bibinfo{address}{Hoboken, NJ}.
\newblock \DOIprefix\doi{10.1002/047174882X}.
%Type = Article
\bibitem[{Cressie and Read(1984)}]{cressieMultinomialGoodnessoffitTests1984}
\bibinfo{author}{Cressie, N.}, \bibinfo{author}{Read, T.R.C.},
  \bibinfo{year}{1984}.
\newblock \bibinfo{title}{Multinomial goodness-of-fit tests}.
\newblock \bibinfo{journal}{Journal of the Royal Statistical Society: Series B}
  \bibinfo{volume}{46}, \bibinfo{pages}{440--464}.
\newblock \DOIprefix\doi{10.1111/j.2517-6161.1984.tb01318.x}.
%Type = Article
\bibitem[{{den Boer} and Sierag(2021)}]{denboerDecisionbasedModelSelection2021}
\bibinfo{author}{{den Boer}, A.V.}, \bibinfo{author}{Sierag, D.D.},
  \bibinfo{year}{2021}.
\newblock \bibinfo{title}{Decision-based model selection}.
\newblock \bibinfo{journal}{European Journal of Operational Research}
  \bibinfo{volume}{290}, \bibinfo{pages}{671--686}.
\newblock \DOIprefix\doi{10.1016/j.ejor.2020.08.025}.
%Type = Inproceedings
\bibitem[{Denisov et~al.(2018)Denisov, Fahland and {van der
  Aalst}}]{denisovUnbiasedFineGrainedDescription2018}
\bibinfo{author}{Denisov, V.}, \bibinfo{author}{Fahland, D.},
  \bibinfo{author}{{van der Aalst}, W.M.P.}, \bibinfo{year}{2018}.
\newblock \bibinfo{title}{Unbiased, {{Fine-Grained Description}} of {{Processes
  Performance}} from {{Event Data}}}, in: \bibinfo{booktitle}{Business
  {{Process Management}} ({{BPM}} 2018)}, \bibinfo{publisher}{Springer}. pp.
  \bibinfo{pages}{139--157}.
\newblock \DOIprefix\doi{10.1007/978-3-319-98648-7_9}.
%Type = Book
\bibitem[{Deza and Deza(2016)}]{dezaEncyclopediaDistances2016}
\bibinfo{author}{Deza, M.M.}, \bibinfo{author}{Deza, E.}, \bibinfo{year}{2016}.
\newblock \bibinfo{title}{Encyclopedia of {{Distances}}}.
\newblock \bibinfo{edition}{4th} ed., \bibinfo{publisher}{Springer},
  \bibinfo{address}{Berlin, Heidelberg}.
\newblock \DOIprefix\doi{10.1007/978-3-662-52844-0}.
%Type = Article
\bibitem[{Esensoy and Carter(2018)}]{Esensoy_2018}
\bibinfo{author}{Esensoy, A.V.}, \bibinfo{author}{Carter, M.W.},
  \bibinfo{year}{2018}.
\newblock \bibinfo{title}{High-fidelity whole-system patient flow modeling to
  assess health care transformation policies}.
\newblock \bibinfo{journal}{European Journal of Operational Research}
  \bibinfo{volume}{266}, \bibinfo{pages}{221--237}.
\newblock \DOIprefix\doi{10.1016/j.ejor.2017.09.019}.
%Type = Inproceedings
\bibitem[{Evermann et~al.(2016)Evermann, Thaler and
  Fettke}]{evermannClusteringTracesUsing2016}
\bibinfo{author}{Evermann, J.}, \bibinfo{author}{Thaler, T.},
  \bibinfo{author}{Fettke, P.}, \bibinfo{year}{2016}.
\newblock \bibinfo{title}{Clustering traces using sequence alignment}, in:
  \bibinfo{booktitle}{Business Process Management Workshops (BPM 2016)},
  \bibinfo{publisher}{Springer}. pp. \bibinfo{pages}{179--190}.
\newblock \DOIprefix\doi{10.1007/978-3-319-42887-1_15}.
%Type = Article
\bibitem[{Fattahi et~al.(2023)Fattahi, Keyvanshokooh, Kannan and
  Govindan}]{fattahiResourcePlanningStrategies2023}
\bibinfo{author}{Fattahi, M.}, \bibinfo{author}{Keyvanshokooh, E.},
  \bibinfo{author}{Kannan, D.}, \bibinfo{author}{Govindan, K.},
  \bibinfo{year}{2023}.
\newblock \bibinfo{title}{Resource planning strategies for healthcare systems
  during a pandemic}.
\newblock \bibinfo{journal}{European Journal of Operational Research}
  \bibinfo{volume}{304}, \bibinfo{pages}{192--206}.
\newblock \DOIprefix\doi{10.1016/j.ejor.2022.01.023}.
%Type = Article
\bibitem[{Fei and Meskens(2013)}]{feiClusteringPatientsTrajectories2013}
\bibinfo{author}{Fei, H.}, \bibinfo{author}{Meskens, N.}, \bibinfo{year}{2013}.
\newblock \bibinfo{title}{Clustering of {{Patients}}' {{Trajectories}} with an
  {{Auto-Stopped Bisecting K-Medoids Algorithm}}}.
\newblock \bibinfo{journal}{Journal of Mathematical Modelling and Algorithms}
  \bibinfo{volume}{12}, \bibinfo{pages}{135--154}.
\newblock \DOIprefix\doi{10.1007/s10852-012-9198-0}.
%Type = Inproceedings
\bibitem[{Gal(2023)}]{galEverythingThereKnow2023}
\bibinfo{author}{Gal, A.}, \bibinfo{year}{2023}.
\newblock \bibinfo{title}{Everything there is to know about stochastically
  known logs}, in: \bibinfo{booktitle}{2023 5th {{International Conference}} on
  {{Process Mining}} ({{ICPM}})}, \bibinfo{publisher}{IEEE}. pp.
  \bibinfo{pages}{xvii--xxiii}.
\newblock \DOIprefix\doi{10.1109/ICPM60904.2023.10271980}.
%Type = Article
\bibitem[{Gallai(1967)}]{gallai_transitiv_1967}
\bibinfo{author}{Gallai, T.}, \bibinfo{year}{1967}.
\newblock \bibinfo{title}{{Transitiv orientierbare Graphen}}.
\newblock \bibinfo{journal}{Acta Mathematica Academiae Scientiarum Hungaricae}
  \bibinfo{volume}{18}, \bibinfo{pages}{25--66}.
\newblock \DOIprefix\doi{10.1007/BF02020961}.
%Type = Article
\bibitem[{{Garc{\'i}a-Ba{\~n}uelos} et~al.(2018){Garc{\'i}a-Ba{\~n}uelos}, {van
  Beest}, Dumas and
  La~Rosa}]{garcia-banuelosCompleteInterpretableConformance2018}
\bibinfo{author}{{Garc{\'i}a-Ba{\~n}uelos}, L.}, \bibinfo{author}{{van Beest},
  N.R.T.P.}, \bibinfo{author}{Dumas, M.}, \bibinfo{author}{La~Rosa, M.},
  \bibinfo{year}{2018}.
\newblock \bibinfo{title}{Complete and {{Interpretable Conformance Checking}}
  of {{Business Processes}}}.
\newblock \bibinfo{journal}{IEEE Transactions on Software Engineering}
  \bibinfo{volume}{44}, \bibinfo{pages}{262--290}.
\newblock \DOIprefix\doi{10.1109/TSE.2017.2668418}.
%Type = Article
\bibitem[{Georgiev et~al.(2025)Georgiev, Fleuriot, Papapanagiotou, McPeake,
  Shenkin and Anand}]{georgievComparingCarePathways2025}
\bibinfo{author}{Georgiev, K.}, \bibinfo{author}{Fleuriot, J.D.},
  \bibinfo{author}{Papapanagiotou, P.}, \bibinfo{author}{McPeake, J.},
  \bibinfo{author}{Shenkin, S.D.}, \bibinfo{author}{Anand, A.},
  \bibinfo{year}{2025}.
\newblock \bibinfo{title}{Comparing care pathways between {{COVID-19}} pandemic
  waves using electronic health records: A process mining case study}.
\newblock \bibinfo{journal}{Journal of Healthcare Informatics Research}
  \bibinfo{volume}{9}, \bibinfo{pages}{41--66}.
\newblock \DOIprefix\doi{10.1007/s41666-024-00181-6}.
%Type = Misc
\bibitem[{Ghahfarokhi et~al.(2021)Ghahfarokhi, Berti and {van der
  Aalst}}]{ghahfarokhiProcessComparisonUsing2021}
\bibinfo{author}{Ghahfarokhi, A.F.}, \bibinfo{author}{Berti, A.},
  \bibinfo{author}{{van der Aalst}, W.M.P.}, \bibinfo{year}{2021}.
\newblock \bibinfo{title}{Process {{Comparison Using Object-Centric Process
  Cubes}}}.
\newblock \bibinfo{howpublished}{Preprint, arXiv:2103.07184}.
%Type = Article
\bibitem[{Gischer(1988)}]{gischerEquationalTheoryPomsets1988}
\bibinfo{author}{Gischer, J.L.}, \bibinfo{year}{1988}.
\newblock \bibinfo{title}{The equational theory of pomsets}.
\newblock \bibinfo{journal}{Theoretical Computer Science} \bibinfo{volume}{61},
  \bibinfo{pages}{199--224}.
\newblock \DOIprefix\doi{10.1016/0304-3975(88)90124-7}.
%Type = Article
\bibitem[{Incerto et~al.(2025)Incerto, Vandin and
  Sarv~Ahrabi}]{incertoStochasticConformanceChecking2025}
\bibinfo{author}{Incerto, E.}, \bibinfo{author}{Vandin, A.},
  \bibinfo{author}{Sarv~Ahrabi, S.}, \bibinfo{year}{2025}.
\newblock \bibinfo{title}{Stochastic conformance checking based on
  variable-length {{Markov}} chains}.
\newblock \bibinfo{journal}{Information Systems} \bibinfo{volume}{133},
  \bibinfo{pages}{102561}.
\newblock \DOIprefix\doi{10.1016/j.is.2025.102561}.
%Type = Article
\bibitem[{Jaynes(1957)}]{jaynesInformationTheoryStatistical1957}
\bibinfo{author}{Jaynes, E.T.}, \bibinfo{year}{1957}.
\newblock \bibinfo{title}{Information {{Theory}} and {{Statistical
  Mechanics}}}.
\newblock \bibinfo{journal}{Physical Review} \bibinfo{volume}{106},
  \bibinfo{pages}{620--630}.
\newblock \DOIprefix\doi{10.1103/PhysRev.106.620}.
%Type = Book
\bibitem[{Kaufman and Rousseeuw(1990)}]{kaufmanFindingGroupsData1990}
\bibinfo{author}{Kaufman, L.}, \bibinfo{author}{Rousseeuw, P.J.},
  \bibinfo{year}{1990}.
\newblock \bibinfo{title}{Finding {{Groups}} in {{Data}}: {{An Introduction}}
  to {{Cluster Analysis}}}.
\newblock \bibinfo{publisher}{John Wiley \& Sons}, \bibinfo{address}{Hoboken,
  NJ}.
\newblock \DOIprefix\doi{10.1002/9780470316801}.
%Type = Book
\bibitem[{Kemeny and Snell(1962)}]{KemenySnell1962}
\bibinfo{author}{Kemeny, J.G.}, \bibinfo{author}{Snell, J.L.},
  \bibinfo{year}{1962}.
\newblock \bibinfo{title}{Mathematical Models in the Social Sciences}.
\newblock \bibinfo{publisher}{Ginn}, \bibinfo{address}{Boston}.
%Type = Article
\bibitem[{Ketkov(2024)}]{ketkovStudyDistributionallyRobust2024}
\bibinfo{author}{Ketkov, S.S.}, \bibinfo{year}{2024}.
\newblock \bibinfo{title}{A study of distributionally robust mixed-integer
  programming with {{Wasserstein}} metric: On the value of incomplete data}.
\newblock \bibinfo{journal}{European Journal of Operational Research}
  \bibinfo{volume}{313}, \bibinfo{pages}{602--615}.
\newblock \DOIprefix\doi{10.1016/j.ejor.2023.10.018}.
%Type = Inproceedings
\bibitem[{Kunze et~al.(2011)Kunze, Weidlich and
  Weske}]{kunzeBehavioralSimilarityProper2011}
\bibinfo{author}{Kunze, M.}, \bibinfo{author}{Weidlich, M.},
  \bibinfo{author}{Weske, M.}, \bibinfo{year}{2011}.
\newblock \bibinfo{title}{Behavioral similarity -- {{A}} proper metric}, in:
  \bibinfo{booktitle}{Business Process Management},
  \bibinfo{publisher}{Springer}. pp. \bibinfo{pages}{166--181}.
\newblock \DOIprefix\doi{10.1007/978-3-642-23059-2_15}.
%Type = Article
\bibitem[{Lee et~al.(2026a)Lee, Ti{\v n}o and
  Styles}]{leeDistanceFunctionStochastic2026}
\bibinfo{author}{Lee, A.R.}, \bibinfo{author}{Ti{\v n}o, P.},
  \bibinfo{author}{Styles, I.B.}, \bibinfo{year}{2026}a.
\newblock \bibinfo{title}{Distance function for stochastic matrices}.
\newblock \bibinfo{journal}{Physical Review E} \bibinfo{volume}{114},
  \bibinfo{pages}{014125}.
\newblock \DOIprefix\doi{10.1103/z34j-grq5}.
%Type = Unpublished
\bibitem[{Lee et~al.(2026b)Lee, Ti{\v n}o and
  Styles}]{leeStringDiagramsProcess2026}
\bibinfo{author}{Lee, A.R.}, \bibinfo{author}{Ti{\v n}o, P.},
  \bibinfo{author}{Styles, I.B.}, \bibinfo{year}{2026}b.
\newblock \bibinfo{title}{String {{Diagrams}} for {{Process Mining}}}.
\newblock \bibinfo{note}{{arXiv} preprint, identifier awaited}.
%Type = Article
\bibitem[{Leemans et~al.(2025)Leemans, Brockhoff, {van der Aalst} and
  Polyvyanyy}]{leemansPartiallyOrderedStochastic2025}
\bibinfo{author}{Leemans, S.J.J.}, \bibinfo{author}{Brockhoff, T.},
  \bibinfo{author}{{van der Aalst}, W.M.P.}, \bibinfo{author}{Polyvyanyy, A.},
  \bibinfo{year}{2025}.
\newblock \bibinfo{title}{Partially ordered stochastic conformance checking}.
\newblock \bibinfo{journal}{Knowledge and Information Systems}
  \bibinfo{volume}{67}, \bibinfo{pages}{2291--2319}.
\newblock \DOIprefix\doi{10.1007/s10115-024-02280-7}.
%Type = Article
\bibitem[{Leemans and
  Fahland(2020)}]{leemansInformationPreservingAbstractions2020}
\bibinfo{author}{Leemans, S.J.J.}, \bibinfo{author}{Fahland, D.},
  \bibinfo{year}{2020}.
\newblock \bibinfo{title}{Information-preserving abstractions of event data in
  process mining}.
\newblock \bibinfo{journal}{Knowledge and Information Systems}
  \bibinfo{volume}{62}, \bibinfo{pages}{1143--1197}.
\newblock \DOIprefix\doi{10.1007/s10115-019-01376-9}.
%Type = Article
\bibitem[{Leemans et~al.(2018)Leemans, Fahland and {van der
  Aalst}}]{leemansScalableProcessDiscovery2018}
\bibinfo{author}{Leemans, S.J.J.}, \bibinfo{author}{Fahland, D.},
  \bibinfo{author}{{van der Aalst}, W.M.P.}, \bibinfo{year}{2018}.
\newblock \bibinfo{title}{Scalable process discovery and conformance checking}.
\newblock \bibinfo{journal}{Software \& Systems Modeling} \bibinfo{volume}{17},
  \bibinfo{pages}{599--631}.
\newblock \DOIprefix\doi{10.1007/s10270-016-0545-x}.
%Type = Article
\bibitem[{Leemans et~al.(2026)Leemans, Lu, {Chapela-Campa}, Di~Ciccio, Cohen,
  Depaire, {Fahrenkrog-Petersen}, Gal, {de Leoni}, Leopold,
  {L{\'o}pez-Pintado}, Mannhardt, Maggi, Martin, Montali, Pegoraro, Polyvyanyy,
  Senderovich, Verboven, Watanabe, Weidlich and {van der
  Werf}}]{leemansTKDEsurvey}
\bibinfo{author}{Leemans, S.J.J.}, \bibinfo{author}{Lu, X.},
  \bibinfo{author}{{Chapela-Campa}, D.}, \bibinfo{author}{Di~Ciccio, C.},
  \bibinfo{author}{Cohen, I.}, \bibinfo{author}{Depaire, B.},
  \bibinfo{author}{{Fahrenkrog-Petersen}, S.}, \bibinfo{author}{Gal, A.},
  \bibinfo{author}{{de Leoni}, M.}, \bibinfo{author}{Leopold, H.},
  \bibinfo{author}{{L{\'o}pez-Pintado}, O.}, \bibinfo{author}{Mannhardt, F.},
  \bibinfo{author}{Maggi, F.M.}, \bibinfo{author}{Martin, N.},
  \bibinfo{author}{Montali, M.}, \bibinfo{author}{Pegoraro, M.},
  \bibinfo{author}{Polyvyanyy, A.}, \bibinfo{author}{Senderovich, A.},
  \bibinfo{author}{Verboven, S.}, \bibinfo{author}{Watanabe, A.},
  \bibinfo{author}{Weidlich, M.}, \bibinfo{author}{{van der Werf}, J.M.E.M.},
  \bibinfo{year}{2026}.
\newblock \bibinfo{title}{Stochastic process mining: {{Characteristics}} and
  challenges}.
\newblock \bibinfo{journal}{IEEE Transactions on Knowledge and Data
  Engineering} \bibinfo{volume}{38}, \bibinfo{pages}{7061--7079}.
\newblock \DOIprefix\doi{10.1109/TKDE.2026.3715402}.
%Type = Article
\bibitem[{Leemans et~al.(2021)Leemans, {van der Aalst}, Brockhoff and
  Polyvyanyy}]{leemansStochasticProcessMining2021}
\bibinfo{author}{Leemans, S.J.J.}, \bibinfo{author}{{van der Aalst}, W.M.P.},
  \bibinfo{author}{Brockhoff, T.}, \bibinfo{author}{Polyvyanyy, A.},
  \bibinfo{year}{2021}.
\newblock \bibinfo{title}{Stochastic process mining: {{Earth}} movers'
  stochastic conformance}.
\newblock \bibinfo{journal}{Information Systems} \bibinfo{volume}{102},
  \bibinfo{pages}{101724}.
\newblock \DOIprefix\doi{10.1016/j.is.2021.101724}.
%Type = Article
\bibitem[{Leemans et~al.(2023)Leemans, {van Zelst} and
  Lu}]{leemansPartialorderbasedProcessMining2023}
\bibinfo{author}{Leemans, S.J.J.}, \bibinfo{author}{{van Zelst}, S.J.},
  \bibinfo{author}{Lu, X.}, \bibinfo{year}{2023}.
\newblock \bibinfo{title}{Partial-order-based process mining: A survey and
  outlook}.
\newblock \bibinfo{journal}{Knowledge and Information Systems}
  \bibinfo{volume}{65}, \bibinfo{pages}{1--29}.
\newblock \DOIprefix\doi{10.1007/s10115-022-01777-3}.
%Type = Incollection
\bibitem[{Li et~al.(2025)Li, Leemans and
  Polyvyanyy}]{liJensenShannonDistance2025}
\bibinfo{author}{Li, T.}, \bibinfo{author}{Leemans, S.J.J.},
  \bibinfo{author}{Polyvyanyy, A.}, \bibinfo{year}{2025}.
\newblock \bibinfo{title}{The {{Jensen}}--{{Shannon}} distance for stochastic
  conformance checking}, in: \bibinfo{booktitle}{Process Mining Workshops}.
  \bibinfo{publisher}{Springer Nature Switzerland}, pp.
  \bibinfo{pages}{70--83}.
\newblock \DOIprefix\doi{10.1007/978-3-031-82225-4_6}.
%Type = Inproceedings
\bibitem[{Liss et~al.(2025)Liss, Mensing and {van der
  Aalst}}]{lissObjectCentricCausalNets2025}
\bibinfo{author}{Liss, L.}, \bibinfo{author}{Mensing, C.},
  \bibinfo{author}{{van der Aalst}, W.M.P.}, \bibinfo{year}{2025}.
\newblock \bibinfo{title}{Object-{{Centric Causal Nets}}}, in:
  \bibinfo{editor}{Krogstie, J.}, \bibinfo{editor}{{Rinderle-Ma}, S.},
  \bibinfo{editor}{Kappel, G.}, \bibinfo{editor}{Proper, H.A.} (Eds.),
  \bibinfo{booktitle}{Advanced {{Information Systems Engineering}}},
  \bibinfo{publisher}{Springer Nature Switzerland}, \bibinfo{address}{Cham}.
  pp. \bibinfo{pages}{94--110}.
\newblock \DOIprefix\doi{10.1007/978-3-031-94571-7_6}.
%Type = Inproceedings
\bibitem[{Lu et~al.(2015)Lu, Fahland and {van der
  Aalst}}]{luConformanceCheckingBased2015}
\bibinfo{author}{Lu, X.}, \bibinfo{author}{Fahland, D.}, \bibinfo{author}{{van
  der Aalst}, W.M.P.}, \bibinfo{year}{2015}.
\newblock \bibinfo{title}{Conformance {{Checking Based}} on {{Partially Ordered
  Event Data}}}, in: \bibinfo{booktitle}{Business {{Process Management
  Workshops}} ({{BPM}} 2014)}, \bibinfo{publisher}{Springer}. pp.
  \bibinfo{pages}{75--88}.
\newblock \DOIprefix\doi{10.1007/978-3-319-15895-2_7}.
%Type = Inproceedings
\bibitem[{Mannila and Meek(2000)}]{mannila2000global}
\bibinfo{author}{Mannila, H.}, \bibinfo{author}{Meek, C.},
  \bibinfo{year}{2000}.
\newblock \bibinfo{title}{Global partial orders from sequential data}, in:
  \bibinfo{booktitle}{Proceedings of the Sixth {{ACM SIGKDD}} International
  Conference on Knowledge Discovery and Data Mining}, \bibinfo{publisher}{ACM},
  \bibinfo{address}{Boston, Massachusetts, USA}. pp. \bibinfo{pages}{161--168}.
\newblock \DOIprefix\doi{10.1145/347090.347122}.
%Type = Inproceedings
\bibitem[{Mazhar et~al.(2023)Mazhar, Tariq, Leemans, Goel, Wynn and
  Staib}]{mazharStochasticAwareComparativeProcess2023}
\bibinfo{author}{Mazhar, T.I.}, \bibinfo{author}{Tariq, A.},
  \bibinfo{author}{Leemans, S.J.J.}, \bibinfo{author}{Goel, K.},
  \bibinfo{author}{Wynn, M.T.}, \bibinfo{author}{Staib, A.},
  \bibinfo{year}{2023}.
\newblock \bibinfo{title}{Stochastic-{{Aware Comparative Process Mining}} in
  {{Healthcare}}}, in: \bibinfo{booktitle}{Business {{Process Management}}
  ({{BPM}} 2023)}, \bibinfo{publisher}{Springer}. pp.
  \bibinfo{pages}{341--358}.
\newblock \DOIprefix\doi{10.1007/978-3-031-41620-0_20}.
%Type = Article
\bibitem[{Morgan and Barton(2025)}]{morganStatisticalProcessControl2025}
\bibinfo{author}{Morgan, L.E.}, \bibinfo{author}{Barton, R.R.},
  \bibinfo{year}{2025}.
\newblock \bibinfo{title}{Statistical process control for queue length
  trajectories using {{Fourier}} analysis}.
\newblock \bibinfo{journal}{European Journal of Operational Research}
  \bibinfo{volume}{325}, \bibinfo{pages}{233--246}.
\newblock \DOIprefix\doi{10.1016/j.ejor.2025.03.013}.
%Type = Inproceedings
\bibitem[{Nguyen et~al.(2018)Nguyen, Dumas, La~Rosa and {ter
  Hofstede}}]{nguyenMultiperspectiveComparisonBusiness2018}
\bibinfo{author}{Nguyen, H.}, \bibinfo{author}{Dumas, M.},
  \bibinfo{author}{La~Rosa, M.}, \bibinfo{author}{{ter Hofstede}, A.H.M.},
  \bibinfo{year}{2018}.
\newblock \bibinfo{title}{Multi-perspective {{Comparison}} of {{Business
  Process Variants Based}} on {{Event Logs}}}, in:
  \bibinfo{booktitle}{Conceptual {{Modeling}} ({{ER}} 2018)},
  \bibinfo{publisher}{Springer}. pp. \bibinfo{pages}{449--459}.
\newblock \DOIprefix\doi{10.1007/978-3-030-00847-5_32}.
%Type = Misc
\bibitem[{Pollard et~al.(2019)Pollard, Johnson, Raffa, Celi, Badawi and
  Mark}]{pollardEICUCollaborativeResearch2019}
\bibinfo{author}{Pollard, T.}, \bibinfo{author}{Johnson, A.},
  \bibinfo{author}{Raffa, J.}, \bibinfo{author}{Celi, L.A.},
  \bibinfo{author}{Badawi, O.}, \bibinfo{author}{Mark, R.},
  \bibinfo{year}{2019}.
\newblock \bibinfo{title}{{{eICU Collaborative Research Database}} (version
  2.0)}.
\newblock \bibinfo{howpublished}{PhysioNet. Dataset}.
\newblock \DOIprefix\doi{10.13026/C2WM1R}.
%Type = Article
\bibitem[{Pollard et~al.(2026)Pollard, Moody, Lehman, Gow, Fernandes, Xie,
  Johnson, Mark and Heldt}]{pollardPhysioNetGlobalPlatform2026}
\bibinfo{author}{Pollard, T.}, \bibinfo{author}{Moody, B.E.},
  \bibinfo{author}{Lehman, L.W.H.}, \bibinfo{author}{Gow, B.J.},
  \bibinfo{author}{Fernandes, C.}, \bibinfo{author}{Xie, C.},
  \bibinfo{author}{Johnson, A.}, \bibinfo{author}{Mark, R.G.},
  \bibinfo{author}{Heldt, T.}, \bibinfo{year}{2026}.
\newblock \bibinfo{title}{{{PhysioNet}} as a global platform for biomedical
  research}.
\newblock \bibinfo{journal}{Nature Health} \bibinfo{volume}{1},
  \bibinfo{pages}{792--795}.
\newblock \DOIprefix\doi{10.1038/s44360-026-00096-z}.
%Type = Article
\bibitem[{Pollard et~al.(2018)Pollard, Johnson, Raffa, Celi, Mark and
  Badawi}]{pollardEICUCollaborativeResearch2018}
\bibinfo{author}{Pollard, T.J.}, \bibinfo{author}{Johnson, A.E.W.},
  \bibinfo{author}{Raffa, J.D.}, \bibinfo{author}{Celi, L.A.},
  \bibinfo{author}{Mark, R.G.}, \bibinfo{author}{Badawi, O.},
  \bibinfo{year}{2018}.
\newblock \bibinfo{title}{The {{eICU Collaborative Research Database}}, a
  freely available multi-center database for critical care research}.
\newblock \bibinfo{journal}{Scientific Data} \bibinfo{volume}{5},
  \bibinfo{pages}{180178}.
\newblock \DOIprefix\doi{10.1038/sdata.2018.178}.
%Type = Article
\bibitem[{Polyvyanyy et~al.(2020)Polyvyanyy, Solti, Weidlich, {Di Ciccio} and
  Mendling}]{polyvyanyyMonotonePrecisionRecall2020}
\bibinfo{author}{Polyvyanyy, A.}, \bibinfo{author}{Solti, A.},
  \bibinfo{author}{Weidlich, M.}, \bibinfo{author}{{Di Ciccio}, C.},
  \bibinfo{author}{Mendling, J.}, \bibinfo{year}{2020}.
\newblock \bibinfo{title}{Monotone {{Precision}} and {{Recall Measures}} for
  {{Comparing Executions}} and {{Specifications}} of {{Dynamic Systems}}}.
\newblock \bibinfo{journal}{ACM Transactions on Software Engineering and
  Methodology} \bibinfo{volume}{29}, \bibinfo{pages}{1--41}.
\newblock \DOIprefix\doi{10.1145/3387909}.
%Type = Article
\bibitem[{Potoniec et~al.(2022)Potoniec, Sroka and
  Pawlak}]{potoniecContinuousDiscoveryCausal2022}
\bibinfo{author}{Potoniec, J.}, \bibinfo{author}{Sroka, D.},
  \bibinfo{author}{Pawlak, T.P.}, \bibinfo{year}{2022}.
\newblock \bibinfo{title}{Continuous discovery of {{Causal}} nets for
  non-stationary business processes using the {{Online Miner}}}.
\newblock \bibinfo{journal}{European Journal of Operational Research}
  \bibinfo{volume}{303}, \bibinfo{pages}{1304--1320}.
\newblock \DOIprefix\doi{10.1016/j.ejor.2022.03.046}.
%Type = Article
\bibitem[{Rousseeuw(1987)}]{rousseeuwSilhouettesGraphicalAid1987}
\bibinfo{author}{Rousseeuw, P.J.}, \bibinfo{year}{1987}.
\newblock \bibinfo{title}{Silhouettes: {{A}} graphical aid to the
  interpretation and validation of cluster analysis}.
\newblock \bibinfo{journal}{Journal of Computational and Applied Mathematics}
  \bibinfo{volume}{20}, \bibinfo{pages}{53--65}.
\newblock \DOIprefix\doi{10.1016/0377-0427(87)90125-7}.
%Type = Inproceedings
\bibitem[{{S{\'a}nchez-Charles} et~al.(2016){S{\'a}nchez-Charles},
  {Munt{\'e}s-Mulero}, Carmona and
  Sol{\'e}}]{sanchezcharlesProcessModelComparison2016}
\bibinfo{author}{{S{\'a}nchez-Charles}, D.},
  \bibinfo{author}{{Munt{\'e}s-Mulero}, V.}, \bibinfo{author}{Carmona, J.},
  \bibinfo{author}{Sol{\'e}, M.}, \bibinfo{year}{2016}.
\newblock \bibinfo{title}{Process {{Model Comparison Based}} on {{Cophenetic
  Distance}}}, in: \bibinfo{booktitle}{Business {{Process Management Forum}}
  ({{BPM}} 2016)}, \bibinfo{publisher}{Springer}. pp.
  \bibinfo{pages}{141--158}.
\newblock \DOIprefix\doi{10.1007/978-3-319-45468-9_9}.
%Type = Article
\bibitem[{Schoknecht et~al.(2018)Schoknecht, Thaler, Fettke, Oberweis and
  Laue}]{schoknechtSimilarityBusinessProcess2018}
\bibinfo{author}{Schoknecht, A.}, \bibinfo{author}{Thaler, T.},
  \bibinfo{author}{Fettke, P.}, \bibinfo{author}{Oberweis, A.},
  \bibinfo{author}{Laue, R.}, \bibinfo{year}{2018}.
\newblock \bibinfo{title}{Similarity of {{Business Process Models}}---{{A
  State-of-the-Art Analysis}}}.
\newblock \bibinfo{journal}{ACM Computing Surveys} \bibinfo{volume}{50},
  \bibinfo{pages}{1--33}.
\newblock \DOIprefix\doi{10.1145/3092694}.
%Type = Misc
\bibitem[{Steeman(2013)}]{steemanBPIChallenge20132013}
\bibinfo{author}{Steeman, W.}, \bibinfo{year}{2013}.
\newblock \bibinfo{title}{{{BPI Challenge}} 2013, incidents}.
\newblock \bibinfo{howpublished}{4TU.ResearchData. Dataset}.
\newblock \DOIprefix\doi{10.4121/uuid:500573e6-accc-4b0c-9576-aa5468b10cee}.
%Type = Article
\bibitem[{Szpilrajn(1930)}]{szpilrajnExtensionOrdrePartiel1930}
\bibinfo{author}{Szpilrajn, E.}, \bibinfo{year}{1930}.
\newblock \bibinfo{title}{{Sur l'extension de l'ordre partiel}}.
\newblock \bibinfo{journal}{Fundamenta Mathematicae} \bibinfo{volume}{16},
  \bibinfo{pages}{386--389}.
\newblock \DOIprefix\doi{10.4064/fm-16-1-386-389}.
%Type = Article
\bibitem[{Topuz et~al.(2024)Topuz, Urban and
  Yildirim}]{topuzMarkovianScoreModel2024}
\bibinfo{author}{Topuz, K.}, \bibinfo{author}{Urban, T.L.},
  \bibinfo{author}{Yildirim, M.B.}, \bibinfo{year}{2024}.
\newblock \bibinfo{title}{A {{Markovian}} score model for evaluating provider
  performance for continuity of care: {{An}} explainable analytics approach}.
\newblock \bibinfo{journal}{European Journal of Operational Research}
  \bibinfo{volume}{317}, \bibinfo{pages}{341--351}.
\newblock \DOIprefix\doi{10.1016/j.ejor.2023.08.039}.
%Type = Article
\bibitem[{Valdes et~al.(1982)Valdes, Tarjan and
  Lawler}]{valdesRecognitionSeriesParallel1982}
\bibinfo{author}{Valdes, J.}, \bibinfo{author}{Tarjan, R.E.},
  \bibinfo{author}{Lawler, E.L.}, \bibinfo{year}{1982}.
\newblock \bibinfo{title}{The {{Recognition}} of {{Series Parallel Digraphs}}}.
\newblock \bibinfo{journal}{SIAM Journal on Computing} \bibinfo{volume}{11},
  \bibinfo{pages}{298--313}.
\newblock \DOIprefix\doi{10.1137/0211023}.
%Type = Article
\bibitem[{Van~Bulck et~al.(2024)Van~Bulck, Goossens, Clarner, Dimitsas,
  Fonseca, {Lamas-Fernandez}, Lester, Pedersen, Phillips and
  Rosati}]{vanbulckWhichAlgorithmSelect2024}
\bibinfo{author}{Van~Bulck, D.}, \bibinfo{author}{Goossens, D.},
  \bibinfo{author}{Clarner, J.P.}, \bibinfo{author}{Dimitsas, A.},
  \bibinfo{author}{Fonseca, G.H.G.}, \bibinfo{author}{{Lamas-Fernandez}, C.},
  \bibinfo{author}{Lester, M.M.}, \bibinfo{author}{Pedersen, J.},
  \bibinfo{author}{Phillips, A.E.}, \bibinfo{author}{Rosati, R.M.},
  \bibinfo{year}{2024}.
\newblock \bibinfo{title}{Which algorithm to select in sports timetabling?}
\newblock \bibinfo{journal}{European Journal of Operational Research}
  \bibinfo{volume}{318}, \bibinfo{pages}{575--591}.
\newblock \DOIprefix\doi{10.1016/j.ejor.2024.06.005}.
%Type = Article
\bibitem[{{van der Aa} et~al.(2020){van der Aa}, Leopold and
  Weidlich}]{vanderaaPartialOrderResolution2020}
\bibinfo{author}{{van der Aa}, H.}, \bibinfo{author}{Leopold, H.},
  \bibinfo{author}{Weidlich, M.}, \bibinfo{year}{2020}.
\newblock \bibinfo{title}{Partial {{Order Resolution}} of {{Event Logs}} for
  {{Process Conformance Checking}}}.
\newblock \bibinfo{journal}{Decision Support Systems} \bibinfo{volume}{136},
  \bibinfo{pages}{113347}.
\newblock \DOIprefix\doi{10.1016/j.dss.2020.113347}.
%Type = Book
\bibitem[{{van der Aalst}(2016)}]{aalstProcessMiningData2016}
\bibinfo{author}{{van der Aalst}, W.M.P.}, \bibinfo{year}{2016}.
\newblock \bibinfo{title}{Process {{Mining}}: {{Data Science}} in {{Action}}}.
\newblock \bibinfo{publisher}{Springer}.
\newblock \DOIprefix\doi{10.1007/978-3-662-49851-4}.
%Type = Article
\bibitem[{{van der Aalst}(2019)}]{vanderaalstPractitionersGuideProcess2019}
\bibinfo{author}{{van der Aalst}, W.M.P.}, \bibinfo{year}{2019}.
\newblock \bibinfo{title}{A practitioner's guide to process mining:
  {{Limitations}} of the directly-follows graph}.
\newblock \bibinfo{journal}{Procedia Computer Science} \bibinfo{volume}{164},
  \bibinfo{pages}{321--328}.
\newblock \DOIprefix\doi{10.1016/j.procs.2019.12.189}.
%Type = Article
\bibitem[{{van der Aalst}(2023)}]{vanderaalstObjectCentricProcessMining2023}
\bibinfo{author}{{van der Aalst}, W.M.P.}, \bibinfo{year}{2023}.
\newblock \bibinfo{title}{Object-{{Centric Process Mining}}: {{Unraveling}} the
  {{Fabric}} of {{Real Processes}}}.
\newblock \bibinfo{journal}{Mathematics} \bibinfo{volume}{11},
  \bibinfo{pages}{2691}.
\newblock \DOIprefix\doi{10.3390/math11122691}.
%Type = Article
\bibitem[{{van der Aalst} and
  Berti(2020)}]{vanderaalstDiscoveringObjectCentricPetri2020}
\bibinfo{author}{{van der Aalst}, W.M.P.}, \bibinfo{author}{Berti, A.},
  \bibinfo{year}{2020}.
\newblock \bibinfo{title}{Discovering {{Object-Centric Petri Nets}}}.
\newblock \bibinfo{journal}{Fundamenta Informaticae} \bibinfo{volume}{175},
  \bibinfo{pages}{1--40}.
\newblock \DOIprefix\doi{10.3233/FI-2020-1946}.
%Type = Book
\bibitem[{{van der Vaart}(1998)}]{vandervaartAsymptoticStatistics1998}
\bibinfo{author}{{van der Vaart}, A.W.}, \bibinfo{year}{1998}.
\newblock \bibinfo{title}{Asymptotic {{Statistics}}}.
\newblock Cambridge {{Series}} in {{Statistical}} and {{Probabilistic
  Mathematics}}, \bibinfo{publisher}{Cambridge University Press},
  \bibinfo{address}{Cambridge}.
\newblock \DOIprefix\doi{10.1017/CBO9780511802256}.
%Type = Misc
\bibitem[{{van Dongen}(2012)}]{vandongenBPIChallenge20122012}
\bibinfo{author}{{van Dongen}, B.F.}, \bibinfo{year}{2012}.
\newblock \bibinfo{title}{{{BPI Challenge}} 2012}.
\newblock \bibinfo{howpublished}{4TU.ResearchData. Dataset}.
\newblock \DOIprefix\doi{10.4121/uuid:3926db30-f712-4394-aebc-75976070e91f}.
%Type = Article
\bibitem[{Walonoski et~al.(2018)Walonoski, Kramer, Nichols, Quina, Moesel,
  Hall, Duffett, Dube, Gallagher and
  McLachlan}]{walonoskiSyntheaApproachMethod2018}
\bibinfo{author}{Walonoski, J.}, \bibinfo{author}{Kramer, M.},
  \bibinfo{author}{Nichols, J.}, \bibinfo{author}{Quina, A.},
  \bibinfo{author}{Moesel, C.}, \bibinfo{author}{Hall, D.},
  \bibinfo{author}{Duffett, C.}, \bibinfo{author}{Dube, K.},
  \bibinfo{author}{Gallagher, T.}, \bibinfo{author}{McLachlan, S.},
  \bibinfo{year}{2018}.
\newblock \bibinfo{title}{Synthea: {{An}} approach, method, and software
  mechanism for generating synthetic patients and the synthetic electronic
  health care record}.
\newblock \bibinfo{journal}{Journal of the American Medical Informatics
  Association} \bibinfo{volume}{25}, \bibinfo{pages}{230--238}.
\newblock \DOIprefix\doi{10.1093/jamia/ocx079}.
%Type = Article
\bibitem[{Wang et~al.(2020)Wang, You, Song and
  Zhang}]{wangWassersteinDistributionallyRobust2020}
\bibinfo{author}{Wang, Z.}, \bibinfo{author}{You, K.}, \bibinfo{author}{Song,
  S.}, \bibinfo{author}{Zhang, Y.}, \bibinfo{year}{2020}.
\newblock \bibinfo{title}{Wasserstein distributionally robust shortest path
  problem}.
\newblock \bibinfo{journal}{European Journal of Operational Research}
  \bibinfo{volume}{284}, \bibinfo{pages}{31--43}.
\newblock \DOIprefix\doi{10.1016/j.ejor.2020.01.009}.
%Type = Article
\bibitem[{Weidlich et~al.(2011)Weidlich, Mendling and
  Weske}]{weidlichEfficientConsistencyMeasurement2011}
\bibinfo{author}{Weidlich, M.}, \bibinfo{author}{Mendling, J.},
  \bibinfo{author}{Weske, M.}, \bibinfo{year}{2011}.
\newblock \bibinfo{title}{Efficient {{Consistency Measurement Based}} on
  {{Behavioral Profiles}} of {{Process Models}}}.
\newblock \bibinfo{journal}{IEEE Transactions on Software Engineering}
  \bibinfo{volume}{37}, \bibinfo{pages}{410--429}.
\newblock \DOIprefix\doi{10.1109/TSE.2010.96}.
%Type = Inproceedings
\bibitem[{Weijters and Ribeiro(2011)}]{weijtersFlexibleHeuristicsMiner2011}
\bibinfo{author}{Weijters, A.J.M.M.}, \bibinfo{author}{Ribeiro, J.T.S.},
  \bibinfo{year}{2011}.
\newblock \bibinfo{title}{Flexible {{Heuristics Miner}} ({{FHM}})}, in:
  \bibinfo{booktitle}{IEEE Symposium on Computational Intelligence and Data
  Mining (CIDM)}, \bibinfo{publisher}{IEEE}. pp. \bibinfo{pages}{310--317}.
\newblock \DOIprefix\doi{10.1109/CIDM.2011.5949453}.
%Type = Article
\bibitem[{Yakowitz and
  Spragins(1968)}]{yakowitzIdentifiabilityFiniteMixtures1968}
\bibinfo{author}{Yakowitz, S.J.}, \bibinfo{author}{Spragins, J.D.},
  \bibinfo{year}{1968}.
\newblock \bibinfo{title}{On the identifiability of finite mixtures}.
\newblock \bibinfo{journal}{The Annals of Mathematical Statistics}
  \bibinfo{volume}{39}, \bibinfo{pages}{209--214}.
\newblock \DOIprefix\doi{10.1214/aoms/1177698520}.

\end{thebibliography}
\end{document}